\documentclass[12pt, a4paper]{article}
\usepackage[T2A]{fontenc}
\usepackage[utf8]{inputenc}
\usepackage{cite}
\usepackage{mathtext}
\usepackage{amssymb,amsthm,amsmath}
\usepackage{array}
\usepackage[english]{babel}
\usepackage[title,titletoc]{appendix}
\usepackage{graphicx}
\usepackage{graphics}
\usepackage{float}

\begin{document}
\author{\bf Yu.A.\,Markov$^{\,1,\,4}\!\,$\thanks{e-mail:markov@icc.ru},\,
M.A.\,Markova$^{\,1}\!\,$\thanks{e-mail:markova@icc.ru},\,
N.Yu.\,Markov$^{\,2}\!\,$\thanks{e-mail:markov@icc.ru},\,
D.M.\,Gitman$^{\,3,\,4,\,5}\!\;$\thanks{e-mail:dmitrygitman@hotmail.com}}

\title{Hamiltonian formalism for Bose excitations in a plasma with a non-Abelian interaction}
%
%
\date{\it\normalsize
\begin{itemize}
\item[]$^{1}$Matrosov Institute of System Dynamics and Control Theory, Siberian Branch, Russian Academy of Sciences, Irkutsk, 664033 Russia
\vspace{-0.3cm}
\item[]$^{2}$Irkutsk State University, Irkutsk, 664003 Russia
\vspace{-0.3cm}
\item[]$^{3}$Lebedev Physical Institute, Russian Academy of Sciences, Moscow, 119991 Russia
\vspace{-0.3cm}
\item[]$^{4}$Tomsk State University, Tomsk, 634050 Russia
\vspace{-0.3cm}
\item[]$^{5}$Institute of Physics, Sao Paulo University, Sao Paulo, 05508-090 Brazil
\end{itemize}}
%
\thispagestyle{empty}
\maketitle{}
\def\theequation{\arabic{section}.\arabic{equation}}
{
\[
\mbox{\bf Аннотация}
\]
We have developed the Hamiltonian theory for collective longitudinally polarized colorless excitations
(plasmons) in a high-temperature gluon plasma using the general formalism for constructing the wave
theory in nonlinear media with dispersion, which was developed by V.\,E.~Zakharov. In this approach, we have
explicitly obtained a special canonical transformation that makes it possible to simplify the Hamiltonian of
interaction of soft gluon excitations and, hence, to derive a new effective Hamiltonian. The approach developed
here is used for constructing a Boltzmann-type kinetic equation describing elastic scattering of collective
longitudinally polarized excitations in a gluon plasma as well as the effect of the so-called nonlinear Landau
damping. We have performed detailed comparison of the effective amplitude of the plasmon–plas\-mon
interaction, which is determined using the classical Hamilton theory, with the corresponding matrix element
calculated in the framework of high-temperature quantum chromodynamics; this has enabled us to determine
applicability limits for the purely classical approach described in this study.
}

{}


\newpage


\section{Introduction}
\setcounter{equation}{0}

It is shown in the theory of usual electron–ion plasma that weak turbulence of the plasma can be of two types (see, for example,  \cite{kadomzev_1964}). Weak turbulence of the first type is caused by scattering of waves by plasma particles. Weak turbulence of the second type is due to decay, fusion, and scattering of waves off one another, which occur without energy exchange between particles and waves. In a number of publications \cite{kovrizhnykh_1965, zakharov_1966, liperovski_1969, zakharov_1985, zakharov_1974, balk_1990}, the kinetic equations for the simplest collective excitations (Langmuir plasmons) of the electron–ion plasma, which describe elastic scattering of plasmons off one another, were constructed and analyzed in detail.\\
\indent At present, a certain interest is shown in the construction of kinetic description of the new fundamental state of matter (quark–gluon plasma that consists of asymptotically free quarks, antiquarks, and gluons; see, for example, review \cite{blaizot_2002}), which is probably formed during ultrarelativistic heavy ion collisions. It is shown that in the high-temperature limit, quark–gluon plasma is successfully described by the effective perturbation theory by Braaten and Pisarski \cite{braaten_1990} reformulated in the terms of the Blaizot-Iancu kinetic equations \cite{blaizot_1994}. A gluon plasma (here, we will disregard for simplicity the presence of quarks and antiquarks) can be represented as a combination of two subsystems, viz., the subsystem of hard thermal gluons and the subsystem of soft plasma excitations, which exchange energy with each other. In a high-temperature gluon plasma, as well as in the usual electron–ion plasma, two types of collective plasma excitations exist, viz., transverse-polarized and longitudinal-polarized excitations (plasmons). In the absence of external chromomagnetic and chromoelectric fields, the color matrix of the number density of collective gluon excitations is diagonal; therefore, these excitations should be treated as colorless.\\
\indent In \cite{markov_2002}, a kinetic description of the nonlinear interaction of colorless and color plasmons in the hard thermal loop approximation \cite{braaten_1990, blaizot_1994} was developed. This approach is based on calculation of some effective currents generating these processes. Using these currents, the matrix elements of nonlinear interaction of an arbitrary (even) number of colorless plasmons \cite{braaten_1990, blaizot_1994} are determined. In this study, we propose an alternative method for kinetic description of the nonlinear plasmon dynamics, which is based on the classical Hamiltonian formalism for systems with distributed parameters and which has been systematically developed by Zakharov \cite{zakharov_1985, zakharov_1974, balk_1990} and Gitman and Tyutin\cite{gitman_1986}. In our case, this approach is based on the fact that equations describing a collisionless high-temperature plasma in the hard thermal loop approximation have the Hamiltonian structure that has been determined in  papers by Nair \cite{nair_1993, nair_1994}, Blaizot и Iancu \cite{blaizot_1995}.  This enables us to develop (at least for weakly excited states; see Conclusions) an independent approach to the derivation of the kinetic equation for soft longitudinally polarized gluonic plasma excitations. In the Hamiltonian approach, the matrix elements of the plasmon–plasmon interaction are obtained using special canonical transformations simplifying the plasmon interaction Hamiltonian.\\
\indent This article has the following structure. In Section 2, we derive the fourth-order effective Hamilton operator $\widetilde{\hat{H}}_{4}$ describing elastic scattering of two colorless plasmons off each other. In Section 3, we introduce plasmon distribution function $N^{l}_{\bf k}$  and analyze the fourth- and sixth-order correlation functions in plasmon creation and annihilation operators $\hat{c}^{\dag\ \!\!a}_{{\bf k}}$ and $\hat{c}^{\phantom{\dagger} \!a}_{{\bf k}}$. Section 4 is devoted to the derivation of the Boltzmann-type kinetic equation for soft gluon excitations with allowance for the nonlinear Landau damping effect for plasmons. Sections 5 and 6 are connected with the determination of the explicit form of three\hspace{0.02cm}- and four-plasmon vertex functions using the hard thermal loop approximation and the approximation of the effective gluon propagator at the plasmon pole. In concluding Section 7, we outline possible ways for generalization of the Hamiltonian description to the case of a strongly excited gluon plasma.\\
\indent In Appendix, we give all basic expressions for the effective gluon vertex functions and gluon propagator in the high-temperature approximation of hard thermal loops.


\section{\bf Colorless plasmon interaction Hamiltonian}
\setcounter{equation}{0}

Let us consider the application of the general Zakharov theory to a specific system (high-temperature gluon plasma) in the semi-classical approximation. The gauge field potentials describing the gluon field in the system are $N_c\times N_c$ matrices in the color space and are defined in terms of $A_{\mu}(x) = A_{\mu}^{a}(x)\, t^{a}$ with $N^{2}_c - 1$ Hermitian generators $t^{a}$ of the color $SU(N_c)$ group in the fundamental representation\footnote{\,The color index $a$ runs through values $1,2,\,\ldots\,,N^{2}_{c}-1$, while vector index $\mu$ runs through values $0,1,2,3$. Everywhere in this article, we imply summation over repeated indices and use the system of units with $\hbar = c = 1$.}. Field strength tensor $F_{\mu\nu}(x) = F^{a}_{\mu\nu}(x)\, t^{a}$, where
\[
F^{a}_{\mu\nu}(x) = \partial^{\phantom{a}}_{\mu} A_{\nu}^{a} -
\partial^{\phantom{a}}_{\nu} A_{\mu}^{a} + g f^{abc} A_{\mu}^{b} A_{\nu}^{c}
\]
obeys the Yang–Mills equation in the $A_0$\hspace{0.03cm}-\hspace{0.03cm}gauge:
\[
\partial_\mu F^{\mu\nu}(x) - ig\hspace{0.03cm}[A_{\mu}(x),F^{\mu\nu}(x)] -
{\xi}_0^{-1} n_{\mu}n^{\nu}A_{\nu}(x) = -j^{\nu}(x),
\]
where $\xi_{0}$ is the gauge parameter in the given gauge. We will henceforth identify the four-vector $n_{\mu}$ with global four-velocity $u_{\mu}$ of the plasma. Color current $j^{\nu}$ is defined conventionally:
\[
j^{\nu}(x) = g\hspace{0.02cm}t^{a}\!\!\int\!d^{4}p \, p^{\nu\,}{\rm Tr}\,(T^{a}f_{g}(x,p)).
\]
Here, $x=(t,{\bf x})$ is the space–time variable of the initial dynamical system and $(T^{a})^{\ \!\!b\ \!\!c}\equiv -if^{\ \!\!a\ \!\!b\ \!\!c}$ is the color matrix in the adjoint representation. Gluon distribution function $f_{g} = f_{g}(x,p)$ is an $(N_{c}^2 - 1)\times (N_{c}^2 - 1)$ Hermitian matrix in the colour space.\\
\indent It is known that there exist two types of physical boson soft (transverse- and longitudinal-polarized) fields in an equilibrium hot quark–gluon plasma \cite{blaizot_2002}. For simplicity, we confine our analysis only to processes involving longitudinally polarized plasma excitations, which are known as {\it plasmons}. These excitations are a purely collective effect of the medium, which has no analogs in the conventional quantum field theory. Let us consider the longitudinal part of the gauge field potential in the form of expansion
$$
\hat{A}^{a}_{\mu}(x) = \int\!\frac{d{\bf k}}{(2\pi)^{3}}\!\left(\frac{Z^{l}({\bf k})}
{2\omega^{l}_{{\bf k}}}\right)^{\!\!1/2}\!\!
\left\{\epsilon^{\ \! l}_{\mu}\ \hat{a}^{\phantom{\dag}\!\!a}_{{\bf k}}\ \!e^{-i\hspace{0.02cm}{\bf k}\cdot{\bf x}}
+
\epsilon^{*l}_{\mu}\ \hat{a}^{\dag\ \!\!a}_{{\bf k}}\ \!e^{i\hspace{0.02cm}{\bf k}\cdot{\bf x}}\right\}, \quad k_{0} 
= \omega^{l}_{{\bf k}},
\eqno{(2.1)}
$$
where $\epsilon^{\ \! l}_{\mu} = \epsilon^{\ \! l}_{\mu} ({\bf k})$ is the polarization vector of a longitudinal plasmon; its explicit form depends on the choice of the gauge (in particular, in the $A_0$\hspace{0.02cm}-\hspace{0.02cm}gauge, this vector is defined by expression (5.6)). Factor $Z^{l}({\bf k})$ is the residue of the effective gluon propagator at the plasmon pole. Coefficients $\hat{a}^{\phantom{\dag}\!\! a}_{{\bf k}}$ and $\hat{a}^{\dag\ \!\! a}_{{\bf k}}$ will be treated as quasiparticle creation and annihilation operators for plasmons obeying the commutation relations for Bose operators:
$$
\Bigl[\hat{a}^{\phantom{\dag}\! a}_{{\bf k}},\,\hat{a}^{\phantom{\dag}\!\! b}_{{\bf k}^{\prime}}\Bigr]
=
\Bigl[\hat{a}^{\dag\ \!\!a}_{{\bf k}},\,\hat{a}^{\dag\ \!\!b}_{{\bf k}^{\prime}}\Bigr] = 0\ ,\ \ \
\Bigl[\hat{a}^{\phantom{\dag}\!\! a}_{{\bf k}},\,\hat{a}^{\dag\ \!\!b}_{{\bf k}^{\prime}}\Bigr]
=
\delta^{\phantom{\dag}\!\! ab}\ \!\! (2\pi)^{3}\hspace{0.03cm}\delta({\bf k} - {\bf k}^{\prime}).
\eqno{(2.2)}
$$
Multiplasmon states are obtained by the multiple action of operator $\hat{a}^{\dag\ \!\!a}_{{\bf k}}$ on vacuum state $|\ \! 0\rangle$, which obeys the following condition:
$$
\hat{a}^{a}_{{\bf k}}|\ \! 0\rangle = 0.
$$
Therefore, we refer as vacuum to the ground unexcited state of the system (i.e., the state without elementary collective excitations). In operators $\hat{a}^{\phantom{\dag}\! a}_{{\bf k}}$ and $\hat{a}^{\dag\ \!\!a}_{{\bf k}}$, only matrix elements corresponding to a change in the number of plasmons by unity differ from zero.\\
\indent Let us write the quantum-mechanical analogue of the Hamilton equation, namely, the Heisenberg equation
for operator $\hat{a}^{a}_{{\bf k}}$:
$$
\frac{\partial\hspace{0.02cm}\hat{a}^{a}_{{\bf k}}}{\partial t}
=
i\hspace{0.03cm}\Bigl[\widehat{H},\hat{a}^{a}_{{\bf k}}\ \!\Bigr].
\eqno{(2.3)}
$$
Here, $\widehat{H}$ is the Hamiltonian of the plasmon system, which is a sum $\widehat{H}=\widehat{H}_{0}+\widehat{H}_{int}$, where
$$
\widehat{H}_{0} = \!\int\!\frac{d{\bf k}}{(2\pi)^{3}}\ \omega^{l}_{{\bf k}}\ \!
\hat{a}^{\dag\ \!\!a}_{{\bf k}}\ \!\hat{a}^{\phantom{\dag}\! a}_{{\bf k}}\
\eqno{(2.4)}
$$
is the Hamiltonian of noninteracting plasmons and $\widehat{H}_{int}$ is interaction Hamiltonian. The dispersion relation $\omega^{\ \! l}_{{\bf k}}$ for plasmons satisfies the following dispersion equation \cite{kalashnikov_1980}:
$$
{\rm Re}\ \!\varepsilon^{l}(\omega,{\bf k})=0\ \!,
\eqno{(2.5)}
$$
where
$$
\varepsilon^{l}(\omega,{\bf k})=1+\frac{3\hspace{0.03cm}\omega^{2}_{pl}}{{\bf k}^{\
\! 2}}\left[1 - F\Biggl(\frac{\omega}{|{\bf k}|^{2}}\Biggr)\right],
\quad
F(x)=\frac{x}{2}\left[\ln\left|\frac{1+x}{1-x} \right|-i\pi\hspace{0.03cm}\theta(1 - |x|)\right]
$$
is the longitudinal permittivity, $\omega^{2}_{pl}=g^{2}N_{c}T^{2}/9$, $T$ is the temperature of the system, and $g$ is the strong interaction constant. In the small amplitude approximation, the interaction Hamiltonian can be written as
a formal integral-power series in $\hat{a}^{a}_{{\bf k}}$ and $\hat{a}^{\dag a}_{{\bf k}}$:
\[
\widehat{H}_{int}=\widehat{H}_{3}+\widehat{H}_{4} +\, \ldots\,\,,
\]
where the third- and fourth-order interaction Hamiltonians have the following structure:
$$
\widehat{H}_{3} = \int\frac{d{\bf k}\, d{\bf k}_{1}\, d{\bf k}_{2}}{(2\pi)^{9}}
\left\{V^{\ a\ a_{1}\ a_{2}}_{{\bf k},\
{\bf k}_{1},\ {\bf k}_{2}}\ \hat{a}^{\dag\ \!a}_{{\bf k}}\
\hat{a}^{\phantom{\dag}\!\!a_{1}}_{{\bf k}_{1}}\
\hat{a}^{\phantom{\dag}\!\!a_{2}}_{{\bf k}_{2}}+V^{*a\ a_{1}\ a_{2}}_{{\bf k},\
{\bf k}_{1},\ {\bf k}_{2}}\ \hat{a}^{\dag\ \!a_{1}}_{{\bf k}_{1}}\
\hat{a}^{\dag\ \!a_{2}}_{{\bf k}_{2}}\ \hat{a}^{\phantom{\dag}\!\!a}_{{\bf k}}
\right\}
\eqno{(2.6)}
$$
\[
\times
(2\pi)^{3}\delta({\bf k} - {\bf k}_{1} - {\bf k}_{2})
\]
\[
+\ \frac{1}{3}\int\frac{d{\bf k}\, d{\bf k}_{1}\, d{\bf k}_{2}}{(2\pi)^{9}}
\left\{U^{\ a\ a_{1}\ a_{2}}_{{\bf k},\
{\bf k}_{1},\ {\bf k}_{2}}\ \hat{a}^{a}_{{\bf k}}\
\hat{a}^{a_{1}}_{{\bf k}_{1}}\
\hat{a}^{a_{2}}_{{\bf k}_{2}}+U^{*a\ a_{1}\ a_{2}}_{{\bf k},\
{\bf k}_{1},\ {\bf k}_{2}}\ \hat{a}^{\dag\ \!a}_{{\bf k}}\
\hat{a}^{\dag\ \!a_{1}}_{{\bf k}_{1}}\ \hat{a}^{\dag\
\!a_{2}}_{{\bf k}_{2}}
\right\}
\]
\[
\times
(2\pi)^{3}\delta({\bf k}+{\bf k}_{1}+{\bf k}_{2}),
\]
$$
\widehat{H}_{4} = \frac{1}{2}\int\frac{d{\bf k}\, d{\bf k}_{1}\, d{\bf k}_{2}\, d{\bf k}_{3}}{(2\pi)^{12}}\ 
T^{\ a\ a_{1}\ a_{2}\ a_{3}}_{{\bf k},\ {\bf k}_{1},\ {\bf k}_{2},\ {\bf k}_{3}}\
\hat{a}^{\dag\ \!a}_{{\bf k}}\ \hat{a}^{\dag\
\!a_{1}}_{{\bf k}_{1}}\ \hat{a}^{\phantom{\dag}\!\!a_{2}}_{{\bf k}_{2}}\
\hat{a}^{\phantom{\dag}\!\!a_{3}}_{{\bf k}_{3}}\
(2\pi)^{3}\delta({\bf k}+{\bf k}_{1}-{\bf k}_{2}-{\bf k}_{3})\eqno{(2.7)}
$$
and so on. Symbol ``$*\,$'' indicates complex conjugation. In expression (2.7), we retained only the ``essential'' contribution in Zakharov’s terminology because the resonance conditions 
$$
\left
\{\begin{array}{l}
{\bf k}+{\bf k}_{1}+{\bf k}_{2}+{\bf k}_{3}=0\\[5pt]
\omega^{l}_{{\bf k}} + \omega^{l}_{{\bf k}_{1}} + \omega^{l}_{{\bf k}_{2}} +
\omega^{l}_{{\bf k}_{3}}=0\ ,
\end{array}\right.\ \ \
\left\{\begin{array}{l}{\bf k}={\bf k}_{1}+{\bf k}_{2}+{\bf k}_{3} \\[5pt]
\omega^{l}_{{\bf k}} = \omega^{l}_{{\bf k}_{1}} + \omega^{l}_{{\bf k}_{2}} +
\omega^{l}_{{\bf k}_{3}}
\end{array}
\right.
$$
have no solutions for the plasmon spectrum defined by dispersion equation (2.5).\\
\indent It should be noted that such a representation of the interaction Hamiltonian in the form of formal infinite power series in the creation and annihilation operators was considered in the monograph by Schwarz \cite{schwarz_1975} based on the quantum field theory for scalar fields.
\par
\begin{figure}[thb]
\centering
\includegraphics*[width=0.95\textwidth]{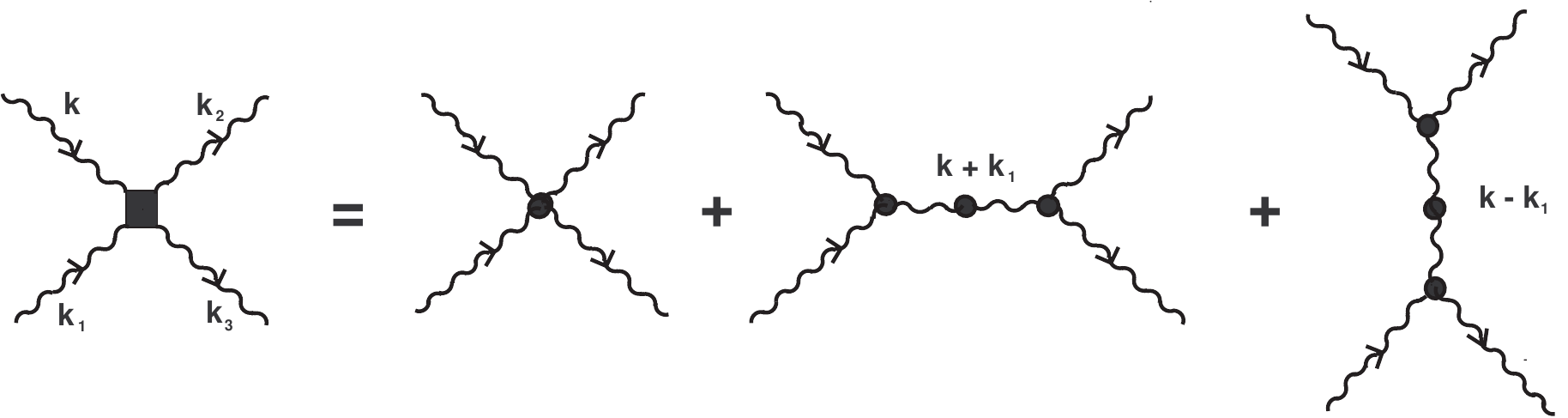}
\caption{Matrix element for the four-plasmon decay. Wavy lines denote plasmons.}
\label{fig1}
\end{figure}

 Coefficients $V^{\ a\ a_{1}\ a_{2}}_{{\bf k},\ {\bf k}_{1},\ {\bf k}_{2}}$,
$U^{\ a\ a_{1}\ a_{2}}_{{\bf k},\ {\bf k}_{1},\ {\bf k}_{2}}$ и
$T^{\ a\ a_{1}\ a_{2}\ a_{3}}_{{\bf k},\ {\bf k}_{1},\ {\bf k}_{2},\ {\bf k}_{3}}$
exhibit certain symmetry:
$$
V^{\ a\ a_{1}\ a_{2}}_{{\bf k},\ {\bf k}_{1},\ {\bf k}_{2}} =
V^{\ a\ a_{2}\ a_{1}}_{{\bf k},\ {\bf k}_{2},\ {\bf k}_{1}}\ ,\ \ \
U^{\ a\ a_{1}\ a_{2}}_{{\bf k},\ {\bf k}_{1},\ {\bf k}_{2}}=U^{\
a\ a_{2}\ a_{1}}_{{\bf k},\ {\bf k}_{2},\ {\bf k}_{1}}=U^{\ a_{1}\
a_{2}\ a}_{{\bf k}_{1},\ {\bf k}_{2},\ {\bf k}}\,,
\eqno{(2.8)}
$$
\vspace{-0.35cm}
$$
T^{\ a\ a_{1}\ a_{2}\ a_{3}}_{{\bf k},\ {\bf k}_{1},\
{\bf k}_{2},\ {\bf k}_{3}}=T^{\ a_{1}\ a\ a_{2}\
a_{3}}_{{\bf k}_{1},\ {\bf k},\ {\bf k}_{2},\ {\bf k}_{3}}=T^{\ a\
a_{1}\ a_{3}\ a_{2}}_{{\bf k},\ {\bf k}_{1},\ {\bf k}_{3},\
{\bf k}_{2}}=T^{*a_{2}\ a_{3}\ a\ a_{1}}_{{\bf k}_{2},\
{\bf k}_{3},\ {\bf k},\ {\bf k}_{1}}.
\hspace{0.25cm}
\eqno{(2.9)}
$$
These coefficient functions determine specific properties of the medium (high-temperature gluon plasma in our case).\\
\indent Let us consider the transformation from operators $\hat{a}^{a}_{{\bf k}}$ to new operators $\hat{c}^{a}_{{\bf k}}$:
$$
\hat{a}^{a}_{{\bf k}} = \hat{c}^{a}_{{\bf k}}\;+
\eqno{(2.10)}
$$
\[
+\!\int\frac{d{\bf k}_{1}\, d{\bf k}_{2}}{(2\pi)^{6}} \left[V^{(1)\ a\ a_{1}\ a_{2}}_{\ \ \ {\bf k},\
{\bf k}_{1},\ {\bf k}_{2}}\ \hat{c}^{a_{1}}_{{\bf k}_{1}}\ \hat{c}^{a_{2}}_{{\bf k}_{2}}
+
V^{(2)\ a\ a_{1}\ a_{2}}_{\ \ \ {\bf k},\ {\bf k}_{1},\ {\bf k}_{2}}\ \hat{c}^{\dag\
\!a_{2}}_{{\bf k}_{2}}\ \hat{c}^{\phantom{\dag}\!\!a_{1}}_{{\bf k}_{1}}
+
V^{(3)\ a\ a_{1}\ a_{2}}_{\ \ \ {\bf k},\ {\bf k}_{1},\ {\bf k}_{2}}\ \hat{c}^{\dag\ \!a_{1}}_{{\bf k}_{1}}\ \hat{c}^{\dag\ \!a_{2}}_{{\bf k}_{2}}\right]
\]
\[
+\!\int\frac{d{\bf k}_{1}\, d{\bf k}_{2}\, d{\bf k}_{3}}{(2\pi)^{9}}
\left[W^{(1)\ a\ a_{1}\ a_{2}\ a_{3}}_{\ \ \ {\bf k},\ {\bf k}_{1},\
{\bf k}_{2},\ {\bf k}_{3}}\ \hat{c}^{a_{1}}_{{\bf k}_{1}}\
\hat{c}^{a_{2}}_{{\bf k}_{2}}\ \hat{c}^{a_{3}}_{{\bf k}_{3}}+\!\ldots\!
+W^{(4)\ a\ a_{1}\ a_{2}\ a_{3}}_{\ \ \ {\bf k},\
{\bf k}_{1},\ {\bf k}_{2},\ {\bf k}_{3}}\ \hat{c}^{\dag\
\!a_{1}}_{{\bf k}_{1}}\ \hat{c}^{\dag\ \!a_{2}}_{{\bf k}_{2}}\
\hat{c}^{\dag\ \!a_{3}}_{{\bf k}_{3}}\right] 
\]
\[
+\,\ldots\,.
\]
The canonicity conditions for this transformation\footnote{\,Variational derivatives with respect to operators $\hat{c}^{\phantom{\dag}\!a}_{\bf k}$ and $\hat{c}^{\dagger\ \!\!a}_{\bf k}$ should be treated as the limits of the corresponding functional derivatives with respect to the classical additions $\varphi^{a}_{\bf k}$ and $\varphi^{*\ \!\!a}_{\bf k}$ to quantum operators $\hat{c}^{\phantom{\dag}\!a}_{\bf k}$ и $\hat{c}^{\dagger\ \!\!a}_{\bf k}$ \cite{bogolyubov_1976}:
\[
\hat{c}^{\phantom{\dag}\!a}_{\bf k} \rightarrow \hat{c}^{\phantom{\dag}\!a}_{\bf k} + \varphi^{\phantom{\dag}\!a}_{\bf k},
\quad
\hat{c}^{\dagger\ \!\!a}_{\bf k} \rightarrow \hat{c}^{\dagger\ \!\!a}_{\bf k} +
\varphi^{*\ \!\!a}_{\bf k}.
\]}
\begin{align}
&\int\! d{\bf k\hspace{0.01cm}}'\!\hspace{0.01cm}\left\{\frac{\delta \hat{a}^{\phantom{\dag}\!\!a}_{{\bf k}}}
{\delta \hat{c}^{\phantom{\dag}\!\!c}_{{\bf k}^{\prime}}}
\,\frac{\delta \hat{a}^{\phantom{\dag}\!\!b}_{{\bf k}''}}
{\delta \hat{c}^{\dagger\ \!\!c}_{{\bf k}^{\prime}}}
\,-\,
\frac{\delta \hat{a}^{\phantom{\dag}\!\!a}_{{\bf k}}}
{\delta \hat{c}^{\dagger\ \!\!c}_{{\bf k}^{\prime}}}\,
\frac{\delta \hat{a}^{\phantom{\dag}\!\!b}_{{\bf k}''}}
{\delta \hat{c}^{\phantom{\dag}\!\!c}_{{\bf k}^{\prime}}}\right\} = 0,
\notag\\[0.8ex]
&\int\! d{\bf k\hspace{0.01cm}}'\!\hspace{0.01cm}\left\{\frac{\delta \hat{a}^{\phantom{\dag}\!\!a}_{{\bf k}}}{\delta \hat{c}^{\phantom{\dag}\!\!c}_{{\bf k}'}}
\,\frac{\delta \hat{a}^{\dagger\ \!\!b}_{{\bf k}''}}{\delta \hat{c}^{\dagger\ \!\!c}_{{\bf k}'}}
\,-\,
\frac{\delta \hat{a}^{\phantom{\dag}\!\!a}_{{\bf k}}}
{\delta \hat{c}^{\dagger\ \!\!c}_{{\bf k}'}}\,
\frac{\delta \hat{a}^{\dagger\ \!\!b}_{{\bf k}''}}
{\delta \hat{c}^{\phantom{\dag}\!\!c}_{{\bf k}'}}\right\} =
\delta^{ab}\delta ({\bf k}-{\bf k}\!\ '')
\notag
\end{align}
impose certain limitations on the coefficient functions of series (2.10). Functions $V^{(1)\ a\ a_{1}\ a_{2}}_{\ \ \ {\bf k},\ {\bf k}_{1},\ {\bf k}_{2}}$, $V^{(2)\ a\ a_{1}\ a_{2}}_{\ \ \ {\bf k},\ {\bf k}_{1},\ {\bf k}_{2}}$ and
$V^{(3)\ a\ a_{1}\ a_{2}}_{\ \ \ {\bf k},\ {\bf k}_{1},\ {\bf k}_{2}}$ must satisfy conditions
$$
V^{(2)\ a\ a_{1}\ a_{2}}_{\ \ \ {\bf k},\ {\bf k}_{1},\ {\bf k}_{2}}
= -2V^{\,*(1)\ a_{1}\ a\ a_{2}}_{\ \ \ {\bf k}_{1},\ {\bf k},\ {\bf k}_{2}}\ ,\
\ \ V^{(3)\ a\ a_{1}\ a_{2}}_{\ \ \ {\bf k},\ {\bf k}_{1},\ {\bf k}_{2}}
= V^{(3)\ a\ a_{2}\ a_{1}}_{\ \ \ {\bf k},\ {\bf k}_{2},\ {\bf k}_{1}}
=V^{(3)\ a_{1}\ a_{2}\ a}_{\ \ \ {\bf k}_{1},\ {\bf k}_{2},\ {\bf k}}\,,
$$
and functions $W^{(i)}_{{\bf k},\ {\bf k}_{1},\ {\bf k}_{2},\ {\bf k}_{3}},\, i = 1,\ldots,4$ satisfy conditions
\[
3\hspace{0.025cm}W^{(1)\ a\ a_{1}\ a_{2}\ a_{3}}_{\ \ \ {\bf k},\ {\bf k}_{1},\ {\bf k}_{2},\ {\bf k}_{3}}
+\,
4\!\int\!\left\{V^{\,*(1)\ a_{2}\ a\ a^{\prime}}_{\ \ \ {\bf k}_{2},\ {\bf k},\ {\bf k}^{\prime}}\
V^{\,*(3)\ a_{1}\ a_{3}\ a^{\prime}}_{\ \ \ {\bf k}_{1},\ {\bf k}_{3},\ {\bf k}^{\prime}}
-
V^{(1)\ a\ a_{2}\ a^{\prime}}_{\ \ \ {\bf k},\ {\bf k}_{2},\ {\bf k}^{\prime}}
V^{(1)\ a^{\prime}\ a_{1}\ a_{3}}_{\ \ \ {\bf k}^{\prime},\ {\bf k}_{1},\ {\bf k}_{3}}\right\}
d{\bf k}
\]
\[
= -\,W^{\,*(3)\ a_{1}\ a\ a_{2}\ a_{3}}_{\ \ \ {\bf k}_{1},\ {\bf k},\ {\bf k}_{2},\ {\bf k}_{3}},
\vspace{0.1cm}
\]
$$
W^{(2)\ a\ a_{1}\ a_{2}\ a_{3}}_{\ \ \ {\bf k},\ {\bf k}_{1},\ {\bf k}_{2},\ {\bf k}_{3}}
+
2\!\int\!\left\{V^{(1)\ a\ a_{1}\ a^{\prime}}_{\ \ \ {\bf k},\ {\bf k}_{1},\ {\bf k}^{\prime}}\
V^{*(1)\ a_{3}\ a_{1}\ a^{\prime}}_{\ \ \ {\bf k}_{3},\ {\bf k}_{2},\ {\bf k}^{\prime}}
+
V^{*(1)\ a^{\prime}\ a\ a_{1}}_{\ \ \ {\bf k}^{\prime},\ {\bf k},\ {\bf k}_{1}}\
V^{(1)\ a^{\prime}\ a_{3}\ a_{2}}_{\ \ \ {\bf k}^{\prime},\ {\bf k}_{3},\ {\bf k}_{2}}
\right.
\hspace{1cm}
$$
$$
\left.-V^{\,*(1)\ a_{1}\ a\ a^{\prime}}_{\ \ \ {\bf k}_{1},\ {\bf k},\ {\bf k}^{\prime}}\
V^{(1)\ a^{\prime}\ a_{3}\ a_{2}}_{\ \ \ {\bf k}^{\prime},\ {\bf k}_{3},\ {\bf k}_{2}}
-
V^{(3)\ a\ a_{1}\ a^{\prime}}_{\ \ \ {\bf k},\ {\bf k}_{1},\ {\bf k}^{\prime}}\
V^{\,*(3)\ a_{3}\ a_{2}\ a^{\prime}}_{\ \ \ {\bf k}_{3},\ {\bf k}_{2},\ {\bf k}^{\prime}}
\right\}d{\bf k}^{\prime}
=
-W^{\,*(2)\ a_{3}\ a_{1}\ a_{2}\ a}_{\ \ \ {\bf k}_{3},\ {\bf k}_{1},\ {\bf k}_{2},\ {\bf k}},
\vspace{0.15cm}
$$
$$
W^{(3)\ a\ a_{1}\ a_{2}\ a_{3}}_{\ \ \ {\bf k},\ {\bf k}_{1},\ {\bf k}_{2},\ {\bf k}_{3}}
+
2\!\int\!\left\{V^{(1)\ a_{3}\ a_{1}\ a^{\prime}}_{\ \ \ {\bf k}_{3},\ {\bf k}_{1},\ {\bf k}^{\prime}}\
V^{(3)\ a\ a_{2}\ a^{\prime}}_{\ \ \ {\bf k},\ {\bf k}_{2},\ {\bf k}^{\prime}}
+
V^{\,*(1)\ a_{1}\ a\ a^{\prime}}_{\ \ \ {\bf k}_{1},\ {\bf k},\ {\bf k}^{\prime}}\
V^{\,*(1)\ a^{\prime}\ a_{3}\ a_{2}}_{\ \ \ {\bf k}^{\prime},\ {\bf k}_{3},\ {\bf k}_{2}}
\right.
\hspace{0.7cm}
$$
$$
\left.
-\,V^{\,*(1)\ a_{1}\ a_{3}\ a^{\prime}}_{\ \ \ {\bf k}_{1},\ {\bf k}_{3},\ {\bf k}^{\prime}}\
V^{*(1)\ a^{\prime}\ a\ a_{2}}_{\ \ \ {\bf k}^{\prime},\ {\bf k},\ {\bf k}_{2}}
-
V^{(1)\ a\ a_{1}\ a^{\prime}}_{\ \ \ {\bf k},\ {\bf k}_{1},\ {\bf k}^{\prime}}\
V^{(3)\ a_{3}\ a_{2}\ a^{\prime}}_{\ \ \ {\bf k}_{3},\ {\bf k}_{2},\ {\bf k}^{\prime}}
\right\}d{\bf k}^{\prime}
=
W^{(3)\ a_{3}\ a_{1}\ a_{2}\ a}_{\ \ \ {\bf k}_{3},\ {\bf k}_{1},\ {\bf k}_{2},\ {\bf k}},
\vspace{0.15cm}
$$
\[
3\hspace{0.03cm}W^{(4)\ a\ a_{1}\ a_{2}\ a_{3}}_{\ \ \ {\bf k},\ {\bf k}_{1},\ {\bf k}_{2},\ {\bf k}_{3}}
+
\,4\!\int\left\{
V^{\,*(1)\ a^{\prime}\ a\ a_{2}}_{\ \ \ {\bf k}^{\prime},\ {\bf k},\ {\bf k}_{2}}\
V^{(3)\ a_{3}\ a_{1}\ a^{\prime}}_{\ \ \ {\bf k}_{3},\ {\bf k}_{1},\ {\bf k}^{\prime}}
-
V^{\,*(1)\ a^{\prime}\ a_{3}\ a_{1}}_{\ \ \ {\bf k}^{\prime},\ {\bf k}_{3},\ {\bf k}_{1}}\
V^{(3)\ a\ a_{2}\ a^{\prime}}_{\ \ \ {\bf k},\ {\bf k}_{2},\ {\bf k}^{\prime}}\right\}
d{\bf k}^{\prime}
\]
\[
=
3\hspace{0.03cm}W^{(4)\ a_{3}\ a_{1}\ a_{2}\ a}_{\ \ \ {\bf k}_{3},\ {\bf k}_{1},\ {\bf k}_{2},\
{\bf k}}.
\]
\indent Because of specific features of dispersion equation (2.5) in a hot gluon plasma, resonance conditions
$$
\left\{
\begin{array}{ll}{\bf k}={\bf k}_{1}+{\bf k}_{2} \\[5pt]
\omega^{l}_{{\bf k}} = \omega^{l}_{{\bf k}_{1}} + \omega^{l}_{{\bf k}_{2}},
\end{array}
\right.
\quad
\left\{
\begin{array}{ll}{\bf k}+{\bf k}_{1}+{\bf k}_{2}=0\\[5pt]
\omega^{l}_{{\bf k}} + \omega^{l}_{{\bf k}_{1}} + \omega^{l}_{{\bf k}_{2}}=0
\end{array}\
\right.
\eqno{(2.11)}
$$
have no solutions (i.e., the longitudinal plasmon spectrum is nondecaying). In this case, canonical transformation (2.10) makes it possible to exclude ``insignificant'' Hamiltonian $\widehat{H}_{3}$ (2.6) by just setting
$$
V^{(1)\ a\ a_{1}\ a_{2}}_{\ \ \ {\bf k},\ {\bf k}_{1},\ {\bf k}_{2}} =
-\frac{V^{\ a\ a_{1}\ a_{2}}_{{\bf k},\ {\bf k}_{1},\ {\bf k}_{2}}}
{\omega^{l}_{{\bf k}} - \omega^{l}_{{\bf k}_{1}} - \omega^{l}_{{\bf k}_{2}}}\
(2\pi)^{3}\delta({\bf k}-{\bf k}_{1}-{\bf k}_{2}),
$$
$$
V^{(3)\ a\ a_{1}\ a_{2}}_{\ \ \ {\bf k},\ {\bf k}_{1},\ {\bf k}_{2}}
= -\frac{U^{*\ a\ a_{1}\ a_{2}}_{{\bf k},\ {\bf k}_{1},\ {\bf k}_{2}}}
{\omega^{l}_{{\bf k}} + \omega^{l}_{{\bf k}_{1}} + \omega^{l}_{{\bf k}_{2}}}\
(2\pi)^{3}\delta({\bf k}+{\bf k}_{1}+{\bf k}_{2}).
$$
This exclusion procedure leads us to following structure of fourth-order effective Hamiltonian $\widetilde{\widehat{H}}_{4}$:
$$
\widetilde{\widehat{H}}_{4} = \frac{1}{2}\int\frac{d{\bf k}\, d{\bf k}_{1}\, d{\bf k}_{2}\, d{\bf k}_{3}}{(2\pi)^{12}}\,
\widetilde{T}^{\ a\ a_{1}\ a_{2}\ a_{3}}_{{\bf k},\ {\bf k}_{1},\
{\bf k}_{2},\ {\bf k}_{3}}\ \hat{c}^{\dag\ \!\!a}_{{\bf k}}\
\hat{c}^{\dag\ \!\!a_{1}}_{{\bf k}_{1}}\
\hat{c}^{\phantom{\dag}\!\!a_{2}}_{{\bf k}_{2}}\
\hat{c}^{\phantom{\dag}\!\!a_{3}}_{{\bf k}_{3}}(2\pi)^{3}\delta({\bf k}+{\bf k}_{1}-{\bf k}_{2}-{\bf k}_{3})\
,\eqno{(2.12)}
$$
where
$$
\widetilde{T}^{\ a\ a_{1}\ a_{2}\ a_{3}}_{{\bf k},\ {\bf k}_{1},\ {\bf k}_{2},\ {\bf k}_{3}}
=
T^{\ a\ a_{1}\ a_{2}\ a_{3}}_{{\bf k},\ {\bf k}_{1},\ {\bf k}_{2},\ {\bf k}_{3}} 
\eqno{(2.13)}
$$
\begin{align}
&-\ \!
2\ \frac{U^{\ b\ a_{2}\ a_{3}}_{-({\bf k}_{2} + {\bf k}_{3}),\ {\bf k}_{2},\ {\bf k}_{3}}\ U^{*\ b\ a\
a_{1}}_{-({\bf k}+{\bf k}_{1}),\ {\bf k},\ {\bf k}_{1}}}{\omega^{l}_{-({\bf k}+{\bf k}_{1})}
+ \omega^{l}_{{\bf k}} + \omega^{l}_{{\bf k}_{1}}}
\ -\ \!2\
\frac{V^{\ b\ a_{2}\ a_{3}}_{{\bf k}_{2}+{\bf k}_{3},\ {\bf k}_{2},\ {\bf k}_{3}}\ V^{*\ b\ a\ a_{1}}_{{\bf k}
+
{\bf k}_{1},\ {\bf k},\ {\bf k}_{1}}}{\omega^{l}_{{\bf k} + {\bf k}_{1}} - \omega^{l}_{{\bf k}}
\omega^{l}_{{\bf k}_{1}}}
\notag\\[0.8ex]
&-2\
\frac{V^{\ a_{1}\ a_{2}\ b}_{{\bf k}_{1},\ {\bf k}_{2},\
{\bf k}_{1}-{\bf k}_{2}}\ V^{*\ a_{3}\ a\ b}_{{\bf k}_{3},\
{\bf k},\ {\bf k}_{3} - {\bf k}}
}{\omega^{l}_{{\bf k}_{3} - {\bf k}} + \omega^{l}_{{\bf k}} - \omega^{l}_{{\bf k}_{3}}}\
-\ \!2\ \!
\frac{V^{\ a\ a_{2}\ b}_{{\bf k},\ {\bf k}_{2},\
{\bf k}-{\bf k}_{2}}\ V^{*\ a_{3}\ a_{1}\ b}_{{\bf k}_{3},\
{\bf k}_{1},\ {\bf k}_{3} - {\bf k}_{1}}
}{\omega^{l}_{{\bf k}_{3} - {\bf k}_{1}} + \omega^{l}_{{\bf k}_{1}} - \omega^{l}_{{\bf k}_{3}}}
\notag\\[0.8ex]
&-2\ \!
\frac{V^{\ a\ a_{3}\ b}_{{\bf k},\ {\bf k}_{3},\
{\bf k}-{\bf k}_{3}}\ V^{*\ a_{2}\ a_{1}\ b}_{{\bf k}_{2},\
{\bf k}_{1},\ {\bf k}_{2} - {\bf k}_{1}}
}{\omega^{l}_{{\bf k}_{2} - {\bf k}_{1}} + \omega^{l}_{{\bf k}_{1}} - \omega^{l}_{{\bf k}_{2}}}
\ -\ \!2\ \!
\frac{V^{\ a_{1}\ a_{3}\ b}_{{\bf k}_{1},\ {\bf k}_{3},\
{\bf k}_{1}-{\bf k}_{3}}\ V^{*\ a_{2}\ a\ b}_{{\bf k}_{2},\
{\bf k},\ {\bf k}_{2}-{\bf k}}
}{\omega^{l}_{{\bf k}_{2} - {\bf k}} + \omega^{l}_{{\bf k}} - \omega^{l}_{{\bf k}_{2}}}\ .
\notag
\end{align}
The determined effective amplitude has a simple diagrammatic interpretation shown in Fig.\,\ref{fig1}. Black square indicates amplitude $\widetilde{T}^{\ a\ a_{1}\ a_{2}\ a_{3}}_{{\bf k},\ {\bf k}_{1},\ {\bf k}_{2},\ {\bf k}_{3}}$.
The first term on the right-hand side of Fig.\,\ref{fig1} defines the direct interaction of four plasmons, which is generated by usual four-plasmon amplitude $T^{\ a\ a_{1}\ a_{2}\ a_{3}}_{{\bf k},\ {\bf k}_{1},\ {\bf k}_{2},\ {\bf k}_{3}}$. The remaining
terms are connected with the interaction generated by amplitudes $U^{\ \! a\ a_{1}\ a_{2}}_{{\bf k},\ {\bf k}_{1},\ {\bf k}_{2}}$ and и $V^{\ \! a\ a_{1}\ a_{2}}_{{\bf k},\ {\bf k}_{1},\ {\bf k}_{2}}$ with intermediate ``virtual'' oscillations. In our case, the conditions of smallness of amplitudes imply that
$$
|\ \!\widetilde{T}^{(4)}| |\ \!c\ \!|^{\,2} \ll\! \Bigl({\bf k}\cdot\partial\omega^{l}_{\bf k}/\partial{\bf k}\Bigr).
\eqno{(2.14)}
$$
\indent Therefore, there exist two equivalent descriptions of the Hamilton system of colorless plasmons for the same physical processes. In the first case, we can use the original Hamiltonian
$$
\widehat{H} = \widehat{H}_{0} + \widehat{H}_{3} + \widehat{H}_{4} +\, \ldots\,,
\eqno{(2.15)}
$$
where $\widehat{H}_0, \widehat{H}_3$, and $\widehat{H}_4$ are defined by expressions (2.4), (2.6), and (2.7), respectively; in the second case, we use Hamiltonian $\tilde{\widehat{H}}$, obtained as a result of the nonlinear
transformation of creation and annihilation Bose
operators $\hat{a}^{\dagger\ \!\!a}_{\bf k}$ and $\hat{a}^{\phantom{\dag}\!\!a}_{\bf k}$:
$$
\widetilde{\widehat{H}} = \widetilde{\widehat{H}}_{0} + \widetilde{\widehat{H}}_{4} +\, \ldots\,,
\hspace{0.7cm}
\eqno{(2.16)}
$$
where, in turn,
$$
\widetilde{\widehat{H}}_{0} = \!\int\!\frac{d{\bf k}}{(2\pi)^{3}}\ \omega^{l}_{{\bf k}}\ \! \hat{c}^{\dag\ \!\!a}_{{\bf k}}\ \!\hat{c}^{\phantom{\dag}\! a}_{{\bf k}},
$$
and operator $\widetilde{\widehat{H}}_4$ is defined by expression (2.12). The Heisenberg equations for operators $\hat{a}^{a}_{\bf k}$ and $\hat{c}^{a}_{\bf k}$ have completely the same form (2.3) with corresponding Hamiltonians (2.15) and (2.16).\\
\indent In connection with this construction, it is appropriate to mention the publication \cite{manko_1997} close to the subject matter of our study, in which a new important concept of nonlinear  $f$-oscillators has been introduced. In \cite{manko_1997} the problem of quantization of a harmonic oscillator was considered, in which the boson creation and annihilation operators were transformed nonlinearly into new creation and annihilation operators determining quantum $f$-oscillators. In this way, a new Hamiltonian with a quite nontrivial structure was obtained; this operator describes the same dynamics as the initial Hamiltonian, as observed in our case.\\
\indent However, despite the closeness of the approaches proposed in the present study and in \cite{manko_1997}, they differ basically. In the approach considered in this section, creation and annihilation operators 
$(\hat{a}^{\dagger\ \!\!a}_{\bf k}, \hat{a}^{\phantom{\dag}\!\!a}_{\bf k})$ and $(\hat{c}^{\dagger\ \!\!a}_{\bf k}, \hat{c}^{\phantom{\dag}\!\!a}_{\bf k})$ and corresponding Hamiltonians (2.15) and (2.16) are connected by the canonical transformation that preserves the standard form of commutation relations (2.2). In the approach described in \cite{manko_1997}, the nonlinear transformations are noncanonical and, hence, the authors modified appropriately commutation relations of type (2.2) for preserving identity of the described dynamics. For this reason, it is impossible in our case to interpret nonlinear oscillations associated with boson operators just as oscillations with a specific energy dependence of the oscillation frequency as in the case of nonlinear $f$-oscillators (however, this fact
may sometimes take place).


\section{\bf Fourth-order correlation function}
\setcounter{equation}{0}

\indent Hamiltonian (2.12) describes elastic scattering of color plasmons by one another (i.e., $2\rightarrow2$ process). The equations of motion for $\hat{c}^{\phantom{\dagger} \!a}_{{\bf k}}$ and $\hat{c}^{\dagger\ \!\!b}_{{\bf k}}$ are determined in this case by the corresponding Heisenberg equations:
\[
\frac{\partial \hat{c}^{\ \!a}_{{\bf k}}}{\partial t}
=
i\Bigl[\widetilde{\widehat{H}}_{0}+\widetilde{\widehat{H}}_{4},\,\hat{c}^{a}_{{\bf k}}\ \!\Bigr]
=
- i\hspace{0.03cm}\omega^{\ \!l}_{{\bf k}}\ \!\hat{c}^{\ \!a}_{{\bf k}}\
\eqno{(3.1)}
\]
\[
-\; i\!\int\frac{d{\bf k}_{1}\, d{\bf k}_{2}\, d{\bf k}_{3}}{(2\pi)^{9}}\ \widetilde{T}^{\ a\ a_{1}\ a_{2}\
a_{3}}_{{\bf k},\ {\bf k}_{1},\ {\bf k}_{2},\ {\bf k}_{3}}\ \hat{c}^{\dag\ \!\!a_{1}}_{{\bf k}_{1}}\
\!\hat{c}^{\phantom{\dag}\!\!a_{2}}_{{\bf k}_{2}}\ \!\hat{c}^{\phantom{\dag}\!\!a_{3}}_{{\bf k}_{3}}\
(2\pi)^{3}\hspace{0.03cm}\delta({\bf k}+{\bf k}_{1}-{\bf k}_{2}-{\bf k}_{3})
\]
and
\[
\frac{\partial \hat{c}^{\dag\ \!\!b}_{{\bf k}}}{\partial t}
=
i\Bigl[\widetilde{\widehat{H}}_{0}+\widetilde{\widehat{H}}_{4},\,\hat{c}^{\dag\ \!\!b}_{\bf k}\ \!\Bigr]
=
i\hspace{0.03cm}\omega^{\ \!l}_{{\bf k}}\ \hat{c}^{\dag\ \!\!b}_{{\bf k}}\
\eqno{(3.2)}
\]
\[
+\; i\!\int\frac{d{\bf k}_{1}\, d{\bf k}_{2}\, d{\bf k}_{3}}{(2\pi)^{9}}\
\widetilde{T}^{\,*\ \!b\ a_{1}\ a_{2}\ a_{3}}_{{\bf k},\ {\bf k}_{1},\ {\bf k}_{2},\ {\bf k}_{3}}\
\hat{c}^{\phantom{\dag}\!\!a_{1}}_{{\bf k}_{1}}\
\!\hat{c}^{\dag\ \!\!a_{2}}_{{\bf k}_{2}}\ \!\hat{c}^{\dag\ \!\!a_{3}}_{{\bf k}_{3}}\
(2\pi)^{3}\hspace{0.03cm}\delta({\bf k}+{\bf k}_{1}-{\bf k}_{2}-{\bf k}_{3}).
\]
These exact equations in the absence of an external color field in the system enable us to determine the kinetic equation for the number density $N^{ab\ \!l}_{{\bf k}\!\!}\! \equiv \delta^{ab} N^{l}_{{\bf k}}$ of colorless plasmons.\\
\indent If the set of waves for a low level of nonlinearity of (2.14) has random phases, this set can be described statistically by introducing correlation function
$$
\langle\,\hat{c}^{\dag\ \!\!a}_{{\bf k}}\ \!\hat{c}^{\phantom{\dag}\!\!b}_{{\bf k}^{\prime}}\rangle
=
\delta^{ab}(2\pi)^{3}\hspace{0.03cm}\delta({\bf k} - {\bf k}^{\prime})\hspace{0.03cm}N^{l}_{{\bf k}}.
\eqno{(3.3)}
$$
It should be emphasized that the introduction of distribution function $N^{l}_{{\bf k}}\equiv N^{l}({\bf k}, {\bf x}, t)$ of quasiparticles (plasmons), which depends on plasmon momentum $\hbar\hspace{0.04cm}{\bf k}$ as well as on coordinate ${\bf x}$ and time $t$, makes sense only in the case when the number of plasmons varies slowly in space and time. This means that the variation of the function over distances on the order of wavelength $\lambda = 2\pi/k$ and in time intervals on the order of wave period $T = 2\pi/\omega^{l}_{\bf k}$ must be much smaller than function $N^{l}_{\bf k}$ itself.\\
\indent Proceeding from Heisenberg equations (3.1) and (3.2), we can determine the kinetic equation for plasmon number density $N^{l}_{{\bf k}}$. For this purpose, we multiply Eqs.\,(3.1) and (3.2) by $\hat{c}^{\dag\ \!\!b}_{{\bf k}^{\prime}}$ and $\hat{c}^{\phantom{\dagger} \!a}_{{\bf k}}$, respectively:
\[
\frac{\partial\hspace{0.02cm}\hat{c}^{\ \!a}_{\bf k}}{\partial\hspace{0.02cm}t}\ \hat{c}^{\dag\ \!\!b}_{{\bf k}^{\prime}}
=
-\, i\hspace{0.03cm}\omega^{\ \!l}_{{\bf k}}\ \!\hat{c}^{\phantom{\dagger} \!a}_{{\bf k}}\ \!
\hat{c}^{\dag\ \!\!b}_{{\bf k}^{\prime}}
\]
\[
-\; i\!\int\frac{d{\bf k}_{1}\, d{\bf k}_{2}\, d{\bf k}_{3}}
{(2\pi)^{9}}\ \widetilde{T}^{\ a\ a_{1}\ a_{2}\ a_{3}}_{{\bf k},\ {\bf k}_{1},\ {\bf k}_{2},\ {\bf k}_{3}}\
\hat{c}^{\dag\ \!\!b}_{{\bf k}}\ \hat{c}^{\dag\ \!\!a_{1}}_{{\bf k}_{1}}\ \!
\!\hat{c}^{\phantom{\dag}\!\!a_{2}}_{{\bf k}_{2}}\ \!\hat{c}^{\phantom{\dag}\!\!a_{3}}_{{\bf k}_{3}}\
(2\pi)^{3}\hspace{0.03cm}\delta({\bf k}+{\bf k}_{1}-{\bf k}_{2}-{\bf k}_{3}),
\]
\[
\hat{c}^{\phantom{\dagger} \!\!a}_{{\bf k}}\
\!\frac{\partial\hspace{0.02cm}\hat{c}^{\dag\ \!\!b}_{{\bf k}^{\prime}}}{\partial\hspace{0.02cm}t}
=
i\hspace{0.03cm}\omega^{\ \!l}_{{\bf k}}\ \! \hat{c}^{\phantom{\dagger} \!\!a}_{{\bf k}}\ \!
\hat{c}^{\dag\ \!\!b}_{{\bf k}^{\prime}}
\]
\[
+\; i\!\int\frac{d{\bf k}_{1}\, d{\bf k}_{2}\, d{\bf k}_{3}}{(2\pi)^{9}}\
\widetilde{T}^{\,*\!\ b\ a_{1}\ a_{2}\ a_{3}}_{{\bf k}^{\prime},\ {\bf k}_{1},\ {\bf k}_{2},\ {\bf k}_{3}}\
\hat{c}^{\phantom{\dagger} \!a}_{{\bf k}}\ \!\hat{c}^{\ \!a_{1}}_{{\bf k}_{1}}\
\!\hat{c}^{\dag\ \!\!a_{2}}_{{\bf k}_{2}}\ \!\hat{c}^{\dag\ \!\!a_{3}}_{{\bf k}_{3}}\
(2\pi)^{3}\hspace{0.03cm}\delta({\bf k}^{\prime} + {\bf k}_{1} - {\bf k}_{2} - {\bf k}_{3}).
\]
Summing these two equations and averaging them, we obtain
\[
\delta^{ab}(2\pi)^{3}\hspace{0.03cm}\delta({\bf k}-{\bf k}\!\ ')
\frac{\partial N^{l}_{{\bf k}}}{\partial t}
=
\eqno{(3.4)}
\]
\[
=
-\, i\!\int\frac{d{\bf k}_{1}\, d{\bf k}_{2}\, d{\bf k}_{3}}{(2\pi)^{9}}\
\biggl\{\widetilde{T}^{\ a\ a_{1}\ a_{2}\ a_{3}}_{{\bf k},\ {\bf k}_{1},\ {\bf k}_{2},\ {\bf k}_{3}}\
I^{\ b\ a_{1}\ a_{2}\ a_{3}}_{{\bf k}^{\prime},\ {\bf k}_{1},\ {\bf k}_{2},\ {\bf k}_{3}}\ \!
(2\pi)^{3}\hspace{0.03cm}\delta({\bf k}+{\bf k}_{1}-{\bf k}_{2}-{\bf k}_{3})
\]
\[
-\ \widetilde{T}^{\,*\!\ b\ a_{1}\ a_{2}\ a_{3}}_{{\bf k}^{\prime},\ {\bf k}_{1},\ {\bf k}_{2},\ {\bf k}_{3}}\
I^{\ a_{2}\ a_{3}\ a\ a_{1}}_{{\bf k}_{2},\ {\bf k}_{3},\ {\bf k},\ {\bf k}_{1}}\ \!
(2\pi)^{3}\hspace{0.03cm}\delta({\bf k}^{\prime}+{\bf k}_{1}-{\bf k}_{2}-{\bf k}_{3})\biggr\},
\]
where
\[
I^{\ a\; a_{1}\; a_{2}\; a_{3}}_{{\bf k},\, {\bf k}_{1},\, {\bf k}_{2},\, {\bf k}_{3}}
=
\langle\ \!\hat{c}^{\dag\ \!\!a}_{{\bf k}}\ \!\hat{c}^{\dag\ \!\!a_{1}}_{{\bf k}_{1}}\ \!
\hat{c}^{\phantom{\dag}\!a_{2}}_{{\bf k}_{2}}\ \!\hat{c}^{\phantom{\dag}\!a_{3}}_{{\bf k}_{3}}\ \!\rangle
\]
is the four-point correlation function. Differentiating the correlation function $I^{\ a\; a_{1}\; a_{2}\; a_{3}}_{{\bf k},\, {\bf k}_{1},\, {\bf k}_{2},\, {\bf k}_{3}}$ with respect to $t$ and considering Eqs.\,(3.1) and (3.2), we obtain the following equation, the right-hand side of which contains sixth-order correlation functions in operators $\hat{c}^{\dag\ \!\!a}_{{\bf k}}$ and $\hat{c}^{\phantom{\dagger} \!a}_{{\bf k}}$:
\[
\frac{\partial I^{\ a\; a_{1}\; a_{2}\; a_{3}}_{{\bf k},\, {\bf k}_{1},\, {\bf k}_{2},\, {\bf k}_{3}}}{\partial t}
=
i\bigl[\ \!\omega^{l}_{{\bf k}} + \omega^{l}_{{\bf k}_{1}} - \omega^{l}_{{\bf k}_{2}} -
\omega^{l}_{{\bf k}_{3}}\bigr]
\ \! I^{\ a\ a_{1}\ a_{2}\ a_{3}}_{{\bf k},\ {\bf k}_{1},\ {\bf k}_{2},\ {\bf k}_{3}}\ +
\eqno{(3.5)}
\]
\[
+\, i\!\int\frac{d{\bf k}^{\prime}_{1}\, d{\bf k}^{\prime}_{2}\, d{\bf k}^{\prime}_{3}}{(2\pi)^{9}}\
\widetilde{T}^{\,*\!\ \!a\ a^{\prime}_{1}\ a^{\prime}_{2}\ a^{\prime}_{3}}_{{\bf k},\ {\bf k}^{\prime}_{1},\ {\bf k}^{\prime}_{2},\ {\bf k}^{\prime}_{3}}\ \!
\langle\ \!\hat{c}^{\ \! a^{\prime}_{1}}_{{\bf k}^{\prime}_{1}}
\ \!\hat{c}^{\dag\ \!\!a^{\prime}_{2}}_{{\bf k}^{\prime}_{2}}
\ \!\hat{c}^{\dag\ \!\!a^{\prime}_{3}}_{{\bf k}^{\prime}_{3}}
\ \!\hat{c}^{\dag\ \!\!a_{1}}_{{\bf k}_{1}}
\ \!\hat{c}^{\phantom{\dag}\!\!a_{2}}_{{\bf k}_{2}}
\ \!\hat{c}^{\phantom{\dag}\!\!a_{3}}_{{\bf k}_{3}}
\ \!\rangle (2\pi)^{3}\hspace{0.03cm}
\delta({\bf k}+{\bf k}^{\prime}_{1}-{\bf k}^{\prime}_{2}-{\bf k}^{\prime}_{3})
\]
\[
+\, i\!\int\frac{d{\bf k}^{\prime}_{1}\, d{\bf k}^{\prime}_{2}\, d{\bf k}^{\prime}_{3}}{(2\pi)^{9}}\
\widetilde{T}^{\,*\!\ \!a_{1}\ a^{\prime}_{1}\ a^{\prime}_{2}\ a^{\prime}_{3}}_{{\bf k}_{1},\
{\bf k}^{\prime}_{1},\ {\bf k}^{\prime}_{2},\ {\bf k}^{\prime}_{3}}\ \!
\langle\ \!\hat{c}^{\dag\ \! a}_{{\bf k}}
\ \!\hat{c}^{a^{\prime}_{1}}_{{\bf k}^{\prime}_{1}}
\ \!\hat{c}^{\dag\ \!\!a^{\prime}_{2}}_{{\bf k}^{\prime}_{2}}
\ \!\hat{c}^{\dag\ \!\!a^{\prime}_{3}}_{{\bf k}^{\prime}_{3}}
\ \!\hat{c}^{\phantom{\dag}\!\!a_{2}}_{{\bf k}_{2}}
\ \!\hat{c}^{\phantom{\dag}\!\!a_{3}}_{{\bf k}_{3}}
\ \!\rangle (2\pi)^{3}\hspace{0.03cm}
\delta({\bf k}_{1} + {\bf k}^{\prime}_{1} - {\bf k}^{\prime}_{2} - {\bf k}^{\prime}_{3})
\]
\[
-\, i\!\int\frac{d{\bf k}^{\prime}_{1}\, d{\bf k}^{\prime}_{2}\, d{\bf k}^{\prime}_{3}}{(2\pi)^{9}}\
\widetilde{T}^{\ \!a_{2}\ a^{\prime}_{1}\ a^{\prime}_{2}\ a^{\prime}_{3}}_{{\bf k}_{2},\
{\bf k}^{\prime}_{1},\ {\bf k}^{\prime}_{2},\ {\bf k}^{\prime}_{3}}\ \!
\langle\ \!\hat{c}^{\dag\ \!\!a}_{{\bf k}}
\ \!\hat{c}^{\dag\ \!\!a_{1}}_{{\bf k}_{1}}
\ \!\hat{c}^{\dag\ \!\!a^{\prime}_{1}}_{{\bf k}^{\prime}_{1}}
\ \!\hat{c}^{\ \! a^{\prime}_{2}}_{{\bf k}^{\prime}_{2}}
\ \!\hat{c}^{\ \! a^{\prime}_{3}}_{{\bf k}^{\prime}_{3}}
\ \!\hat{c}^{\phantom{\dag}\!\!a_{3}}_{{\bf k}_{3}}
\ \!\rangle (2\pi)^{3}\hspace{0.03cm}
\delta({\bf k}_{2}+{\bf k}^{\prime}_{1}-{\bf k}^{\prime}_{2}-{\bf k}^{\prime}_{3})
\]
\[
-\, i\!\int\frac{d{\bf k}^{\prime}_{1}\, d{\bf k}^{\prime}_{2}\, d{\bf k}^{\prime}_{3}}{(2\pi)^{9}}\
\widetilde{T}^{\ \!a_{3}\ a^{\prime}_{1}\ a^{\prime}_{2}\ a^{\prime}_{3}}_{{\bf k}_{3},\
{\bf k}^{\prime}_{1},\ {\bf k}^{\prime}_{2},\ {\bf k}^{\prime}_{3}}\ \!
\langle\ \!\hat{c}^{\dag\ \!\!a}_{{\bf k}}
\ \!\hat{c}^{\dag\ \!\!a_{1}}_{{\bf k}_{1}}
\ \!\hat{c}^{\ \!a_{2}}_{{\bf k}_{2}}
\ \!\hat{c}^{\dag \ \!\!a^{\prime}_{1}}_{{\bf k}^{\prime}_{1}}
\ \!\hat{c}^{\ \! a^{\prime}_{2}}_{{\bf k}^{\prime}_{2}}
\ \!\hat{c}^{\ \! a^{\prime}_{3}}_{{\bf k}^{\prime}_{3}}
\ \!\rangle (2\pi)^{3}\hspace{0.03cm}
\delta({\bf k}_{3} + {\bf k}^{\prime}_{1} - {\bf k}^{\prime}_{2} - {\bf k}^{\prime}_{3}).
\]
We close this set of equations for correlation functions by expressing the formulas for sixth-order correlation functions in terms of pair correlation functions. For example, the first sixth-order correlation function on the right-hand side of Eq.\,(3.5) has the following structure:
\[
\langle\ \!\hat{c}^{\ \!\!a^{\prime}_{1}}_{{\bf k}^{\prime}_{1}}
\ \!\hat{c}^{\dag\ \!\!a^{\prime}_{2}}_{{\bf k}^{\prime}_{2}}
\ \!\hat{c}^{\dag\ \!\!a^{\prime}_{3}}_{{\bf k}^{\prime}_{3}}
\ \!\hat{c}^{\dag\ \!a_{1}}_{{\bf k}_{1}}
\ \!\hat{c}^{\phantom{\dag}\!\!a_{2}}_{{\bf k}_{2}}
\ \!\hat{c}^{\phantom{\dag}\!\!a_{3}}_{{\bf k}_{3}}
\ \!\rangle
=
\eqno{(3.6)}
\]
\[
=
3\hspace{0.03cm}(2\pi)^{9}\Bigl\{
\delta^{\ \!\!a^{\phantom{\prime}}_{3}a^{\prime}_{3}}
\delta^{\ \!\!a^{\phantom{\prime}}_{2}a^{\prime}_{2}}
\delta^{\ \!\!a^{\phantom{\prime}}_{1}a^{\prime}_{1}}\ \!
\delta({\bf k}^{\prime}_{3}-{\bf k}^{\phantom{\prime}}_{3})
\delta({\bf k}^{\prime}_{2}-{\bf k}^{\phantom{\prime}}_{2})
\delta({\bf k}^{\prime}_{1}-{\bf k}^{\phantom{\prime}}_{1})\ \!
N^{l}_{{\bf k}_{3}} N^{l}_{{\bf k}_{2}} N^{l}_{{\bf k}_{1}}
\]
\[
+\
\delta^{\ \!\!a^{\phantom{\prime}}_{3}a^{\prime}_{3}}
\delta^{\ \!\!a_{1}a_{2}}
\delta^{\ \!\!a^{\prime}_{2}a^{\prime}_{1}}\ \!
\delta({\bf k}^{\prime}_{3}-{\bf k}^{\phantom{\prime}}_{3})
\delta({\bf k}_{1}-{\bf k}_{2})
\delta({\bf k}^{\prime}_{2}-{\bf k}^{\prime}_{1})\ \!
N^{l}_{{\bf k}_{3}} N^{l}_{{\bf k}_{1}} N^{l}_{{\bf k}^{\prime}_{1}}
\]
\[
+\
\delta^{\ \!\!a^{\prime}_{3}a^{\prime}_{1}}
\delta^{\ \!\!a_{1}a_{3}}
\delta^{\ \!\!a^{\prime}_{2}a^{\phantom{\prime}}_{2}}\ \!
\delta({\bf k}^{\prime}_{3}-{\bf k}^{\prime}_{1})
\delta({\bf k}_{1}-{\bf k}_{3})
\delta({\bf k}^{\prime}_{2}-{\bf k}^{\phantom{\prime}}_{2})\ \!
N^{l}_{{\bf k}^{\prime}_{1}} N^{l}_{{\bf k}_{1}} N^{l}_{{\bf k}_{2}}
\]
\[
+\
\delta^{\ \!\!a^{\prime}_{3}a^{\prime}_{1}}
\delta^{\ \!\!a_{1}a_{2}}
\delta^{\ \!\!a^{\prime}_{2}a^{\phantom{\prime}}_{3}}\ \!
\delta({\bf k}^{\prime}_{3}-{\bf k}^{\prime}_{1})
\delta({\bf k}_{1}-{\bf k}_{2})
\delta({\bf k}^{\prime}_{2}-{\bf k}^{\phantom{\prime}}_{3})\ \!
N^{l}_{{\bf k}^{\prime}_{1}} N^{l}_{{\bf k}_{1}} N^{l}_{{\bf k}_{3}}
\]
\[
+\
\delta^{\ \!\!a^{\prime}_{3}a^{\phantom{\prime}}_{2}}
\delta^{\ \!\!a_{1}a_{3}}
\delta^{\ \!\!a^{\prime}_{2}a^{\prime}_{1}}\ \!
\delta({\bf k}^{\prime}_{3}-{\bf k}^{\phantom{\prime}}_{2})
\delta({\bf k}_{1}-{\bf k}_{3})
\delta({\bf k}^{\prime}_{2}-{\bf k}^{\prime}_{1})\ \!
N^{l}_{{\bf k}_{2}} N^{l}_{{\bf k}_{1}} N^{l}_{{\bf k}^{\prime}_{1}}
\]
\[
\hspace{0.45cm}
+\
\delta^{\ \!\!a^{\prime}_{3}a^{\phantom{\prime}}_{2}}
\delta^{\ \!\!a^{\prime}_{1}a^{\phantom{\prime}}_{1}}
\delta^{\ \!\!a^{\prime}_{2}a^{\phantom{\prime}}_{3}}\ \!
\delta({\bf k}^{\prime}_{3}-{\bf k}^{\phantom{\prime}}_{2})
\delta({\bf k}^{\prime}_{1}-{\bf k}^{\phantom{\prime}}_{1})
\delta({\bf k}^{\prime}_{2}-{\bf k}^{\phantom{\prime}}_{3})\ \!
N^{l}_{{\bf k}_{2}} N^{l}_{{\bf k}_{1}} N^{l}_{{\bf k}_{3}}\Bigr\}.
\]
In this expression, only the first and last terms make the required contribution to the required kinetic equation. Substituting these terms into the first integral on the right-hand side of Eq.\,(3.5) and performing summation over color indices $a^{\prime}_{1},\, a^{\prime}_{2}$ and $a^{\prime}_{3}$ and integration over momenta ${\bf k}^{\prime}_{1},\, {\bf k}^{\prime}_{2}$ и ${\bf k}^{\prime}_{3}$, we obtain
\[
3\hspace{0.03cm}i\biggl\{\widetilde{T}^{\,*\ \!a\ a_{1}\ a_{2}\ a_{3}}_{{\bf k},\ {\bf k}_{1},\ {\bf k}_{2},\ {\bf k}_{3}}\ \!
N^{l}_{{\bf k}_{1}} N^{l}_{{\bf k}_{2}} N^{l}_{{\bf k}_{3}}
+
\widetilde{T}^{*\ \!a\ a_{1}\ a_{2}\ a_{3}}_{{\bf k},\ {\bf k}_{1},\ {\bf k}_{2},\ {\bf k}_{3}}\ \!
N^{l}_{{\bf k}_{1}} N^{l}_{{\bf k}_{2}} N^{l}_{{\bf k}_{3}}
\biggr\}
\ \!(2\pi)^{3}\hspace{0.03cm}
\delta({\bf k}+{\bf k}_{1}-{\bf k}_{2}-{\bf k}_{3})
\]
\[
= 6\hspace{0.03cm}i\ \!\widetilde{T}^{\,*\ \!a\ a_{1}\ a_{2}\ a_{3}}_{{\bf k},\ {\bf k}_{1},\ {\bf k}_{2},\ {\bf k}_{3}}\ \! N^{l}_{{\bf k}_{1}} N^{l}_{{\bf k}_{2}} N^{l}_{{\bf k}_{3}}
\ \!(2\pi)^{3}\hspace{0.03cm}
\delta({\bf k}+{\bf k}_{1}-{\bf k}_{2}-{\bf k}_{3}).
\eqno{(3.7)}
\]
By way of example, let us consider the explicit form of the contribution that generates the second term in the decomposition of correlation function (3.6):
\[
N^{l}_{\bf k} N^{l}_{{\bf k}_{1}}\delta({\bf k}_{2} - {\bf k}_{1})\ \! \delta({\bf k}_{3} - {\bf k})\ \!
\delta^{a_{1}a_{2}}\ \! i\!\int\!
\widetilde{T}^{\,*\ \!a\ b\ b\ a_{3}}_{{\bf k},\ {\bf k}^{\prime},\ {\bf k}^{\prime},\ {\bf k}}\ \!
N^{l}_{{\bf k}^{\prime}}\ \! d{\bf k}^{\prime}.
\]
Comparing the last two expressions, we see that they have quite different structures.\\
\indent Let us now consider the second six-point correlation function in expression (3.5). In this correlator, we write in explicit form only “regular” terms:
\[
\langle\ \!\hat{c}^{\dag\ \!\!a}_{{\bf k}}
\ \!\hat{c}^{a^{\prime}_{1}}_{{\bf k}^{\prime}_{1}}
\ \!\hat{c}^{\dag\ \!\!a^{\prime}_{2}}_{{\bf k}^{\prime}_{2}}
\ \!\hat{c}^{\dag\ \!\!a^{\prime}_{3}}_{{\bf k}^{\prime}_{3}}
\ \!\hat{c}^{\phantom{\dag}\!\!a_{2}}_{{\bf k}_{2}}
\ \!\hat{c}^{\phantom{\dag}\!\!a_{3}}_{{\bf k}_{3}}
\ \!\rangle
=
\]
\[
=
3\hspace{0.03cm}(2\pi)^{9}\Bigl\{
\delta^{\ \!\!a^{\phantom{\prime}}\!a^{\prime}_{1}}
\delta^{\ \!\!a^{\phantom{\prime}}_{2}a^{\prime}_{2}}
\delta^{\ \!\!a^{\phantom{\prime}}_{3}a^{\prime}_{3}}\ \!
\delta({\bf k}^{\prime}_{1}-{\bf k}^{\phantom{\prime}})
\delta({\bf k}^{\prime}_{2}-{\bf k}^{\phantom{\prime}}_{2})
\delta({\bf k}^{\prime}_{3}-{\bf k}^{\phantom{\prime}}_{3})\ \!
N^{l}_{{\bf k}} N^{l}_{{\bf k}_{2}} N^{l}_{{\bf k}_{3}}
+\ (2\rightleftarrows 3)\ +\ \ldots\,\Bigr\}.
\]
Substituting this expression into the second integral in (3.5), we obtain the following expression analogous to
(3.7):
\[
 6\hspace{0.03cm}i\
 \!\widetilde{T}^{\,*\ \!a_{1}\ a\ a_{2}\ a_{3}}_{{\bf k}_{1},\ {\bf k},\ {\bf k}_{2},\ {\bf k}_{3}}\,
 N^{l}_{{\bf k}} N^{l}_{{\bf k}_{2}} N^{l}_{{\bf k}_{3}}
\ \!(2\pi)^{3}\hspace{0.03cm} \delta({\bf k}+{\bf k}_{1}-{\bf k}_{2}-{\bf k}_{3}).
\]
Similar arguments for the third and fourth correlators in expression (3.5) give the remaining two contributions:
\[
 -6\hspace{0.03cm}i\
\!\widetilde{T}^{\ \!a_{2}\ a_{3}\ a\ a_{1}}_{{\bf k}_{2},\ {\bf k}_{3},\ {\bf k},\ {\bf k}_{1}}\,
N^{l}_{{\bf k}} N^{l}_{{\bf k}_{1}} N^{l}_{{\bf k}_{3}}
\ \!(2\pi)^{3}\hspace{0.03cm} \delta({\bf k}+{\bf k}_{1}-{\bf k}_{2}-{\bf k}_{3})
\]
and
\[
 -6\hspace{0.03cm}i\
\!\widetilde{T}^{\ \!a_{3}\ a_{2}\ a\ a_{1}}_{{\bf k}_{3},\ {\bf k}_{2},\ {\bf k},\ {\bf k}_{1}}\,
N^{l}_{{\bf k}} N^{l}_{{\bf k}_{1}} N^{l}_{{\bf k}_{2}}
\ \!(2\pi)^{3}\hspace{0.03cm} \delta({\bf k}+{\bf k}_{1}-{\bf k}_{2}-{\bf k}_{3}).
\]
Considering the symmetry relations for the scattering amplitude
\[
\widetilde{T}^{\ \!a_{2}\ a_{3}\ a\ a_{1}}_{{\bf k}_{2},\ {\bf k}_{3},\ {\bf k},\ {\bf k}_{1}}
=
\widetilde{T}^{*\ \!a\ a_{1}\ a_{2}\ a_{3}}_{{\bf k},\ {\bf k}_{1},\ {\bf k}_{2},\ {\bf k}_{3}},
\quad
\widetilde{T}^{\ \!a_{3}\ a_{2}\ a\ a_{1}}_{{\bf k}_{3},\ {\bf k}_{2},\ {\bf k},\ {\bf k}_{1}}
=
\widetilde{T}^{*\ \!a\ a_{1}\ a_{2}\ a_{3}}_{{\bf k},\ {\bf k}_{1},\ {\bf k}_{2},\ {\bf k}_{3}},
\]
we obtain the following equation for the fourth-order correlation function, instead of (3.5):
\[
\frac{\partial I^{\ a\ a_{1}\ a_{2}\ a_{3}}_{{\bf k},\ {\bf k}_{1},\ {\bf k}_{2},\ {\bf k}_{3}}}{\partial t}
=
i\bigl[\ \!\omega^{l}_{{\bf k}} + \omega^{l}_{{\bf k}_{1}} - \omega^{l}_{{\bf k}_{2}} -
\omega^{l}_{{\bf k}_{3}}\bigr]
\ \! I^{\ a\; a_{1}\; a_{2}\; a_{3}}_{{\bf k},\, {\bf k}_{1},\, {\bf k}_{2},\, {\bf k}_{3}}
\eqno{(3.8)}
\]
\[
+\; 6\hspace{0.04cm}i\ \!\widetilde{T}^{\,*\ \!a\ a_{1}\ a_{2}\ a_{3}}_{{\bf k},\ {\bf k}_{1},\ {\bf k}_{2},\ {\bf k}_{3}}\ \! \Bigl(
N^{l}_{\bf k} N^{l}_{{\bf k}_{2}} N^{l}_{{\bf k}_{3}} +
N^{l}_{{\bf k}_{1}} N^{l}_{{\bf k}_{2}} N^{l}_{{\bf k}_{3}} -
N^{l}_{{\bf k}} N^{l}_{{\bf k}_{1}} N^{l}_{{\bf k}_{3}} -
N^{l}_{{\bf k}} N^{l}_{{\bf k}_{1}} N^{l}_{{\bf k}_{2}}\Bigr)
\]
\[
\times
\ \!(2\pi)^{3}\hspace{0.03cm}
\delta({\bf k}+{\bf k}_{1}-{\bf k}_{2}-{\bf k}_{3}).
\]


\section{\bf Kinetic equation for gluon excitations}
\setcounter{equation}{0}

Let us now pass to the direct derivation of the kinetic equation for plasmons. The self-consistent set of equations (3.4) and (3.8) determines, in principle, the evolution of plasmon number density $N^{l}_{\bf k}$. However, we introduce one more simplification: in Eq.\,(3.8), we disregard the term with the time derivative as compared to the term containing the difference in the eigenfrequencies of wave packets. Instead of relation (3.8), we have
\[
I^{\ a\; a_{1}\; a_{2}\; a_{3}}_{{\bf k},\, {\bf k}_{1},\, {\bf k}_{2},\, {\bf k}_{3}} \simeq
N^{l}_{{\bf k}} N^{l}_{{\bf k}_{1}}\ \!(2\pi)^{6}\hspace{0.03cm}
\Bigl[\ \! \delta^{a\ \!\!a_{2}} \delta^{a_{1}\ \!\!a_{3}}\ \! \delta({\bf k} - {\bf k}_{2}) \delta({\bf k}_{1} - {\bf k}_{3})
+
\delta^{a\ \!\!a_{3}} \delta^{a_{1}\ \!\!a_{2}}\ \! \delta({\bf k} - {\bf k}_{3}) \delta({\bf k}_{1} - {\bf k}_{2})
\Bigr]
\]
\[
-\; \frac{6}{\Delta\hspace{0.02cm}\omega + i\hspace{0.03cm}0}\
\widetilde{T}^{\,*\ \!a\ a_{1}\ a_{2}\ a_{3}}_{{\bf k},\ {\bf k}_{1},\ {\bf k}_{2},\ {\bf k}_{3}}\ \!
\Bigl(
N^{l}_{\bf k} N^{l}_{{\bf k}_{2}} N^{l}_{{\bf k}_{3}} +
N^{l}_{{\bf k}_{1}} N^{l}_{{\bf k}_{2}} N^{l}_{{\bf k}_{3}} -
N^{l}_{{\bf k}} N^{l}_{{\bf k}_{1}} N^{l}_{{\bf k}_{3}} -
N^{l}_{{\bf k}} N^{l}_{{\bf k}_{1}} N^{l}_{{\bf k}_{2}}\Bigr)
\times
\]
\[
\times
\ \!(2\pi)^{3}\hspace{0.03cm}\delta({\bf k}+{\bf k}_{1}-{\bf k}_{2}-{\bf k}_{3}),
\]
where
\[
\Delta\hspace{0.02cm}\omega \equiv
\omega^{l}_{{\bf k}} + \omega^{l}_{{\bf k}_{1}} - \omega^{l}_{{\bf k}_{2}} - \omega^{l}_{{\bf k}_{3}}.
\]
Here, the first term on the right-hand side, which corresponds to completely uncorrelated waves (purely Gaussian fluctuations) is the solution to the homogeneous equation for fourth-order correlation function
$I^{\ a\; a_{1}\; a_{2}\; a_{3}}_{{\bf k},\, {\bf k}_{1},\, {\bf k}_{2},\, {\bf k}_{3}}$. The second term determines the deviation of the four-point correlator from the Gaussian approximation for a low nonlinearity level of interacting waves.\\
\indent We substitute the first term into the right-hand side of Eq.\,(3.4) for $N^{l}_{\bf k}$:
\[
-\, i\ \!(2\pi)^{3}\hspace{0.03cm}N^{l}_{\bf k}\!\int\!\frac{d{\bf k}_{1}}{(2\pi)^{3}}\
N^{l}_{{\bf k}_{1}}
\Bigl\{\widetilde{T}^{\ \!a\ a_{1}\ b\ a_{1}}_{{\bf k},\ {\bf k}_{1},\ {\bf k}^{\prime},\ {\bf k}_{1}}\ \!
\delta({\bf k} - {\bf k}^{\prime})
\,+\,
\widetilde{T}^{\ \!a\ a_{1}\ a_{1}\ b}_{{\bf k},\ {\bf k}_{1},\ {\bf k}_{1}\ \ {\bf k}^{\prime}}\ \!
\delta({\bf k} - {\bf k}^{\prime})
\]
\[
-\
\widetilde{T}^{\,*\ \!b\ a_{1}\ a\ a_{1}}_{{\bf k}^{\prime},\ {\bf k}_{1},\ {\bf k},\ {\bf k}_{1}}\ \!
\delta({\bf k}^{\prime} - {\bf k})
\,-\,
\widetilde{T}^{\,*\ \!b\ a_{1}\ a_{1}\ a}_{{\bf k}^{\prime},\ {\bf k}_{1},\ {\bf k}_{1},\ {\bf k}}\ \!
\delta({\bf k}^{\prime} - {\bf k})\Bigr\}
\]
\[
= -\, i\ \!2\hspace{0.03cm}(2\pi)^{3}\hspace{0.03cm}\delta({\bf k} - {\bf k}^{\prime})
N^{l}_{\bf k}\!\int\!\frac{d{\bf k}_{1}}{(2\pi)^{3}}\, N^{l}_{{\bf k}_{1}}
\Bigl\{\widetilde{T}^{\ a\ a_{1}\ b\ a_{1}}_{{\bf k},\ {\bf k}_{1},\ {\bf k},\ {\bf k}_{1}}
-
\widetilde{T}^{\,*\ \!b\ a_{1}\ a\ a_{1}}_{{\bf k},\ {\bf k}_{1},\ {\bf k},\ {\bf k}_{1}}\Bigr\}.
\eqno{(4.1)}
\]
Further, we substitute the second term into the right-hand side of Eq.\,(3.4):
\[
-\ \! 6\hspace{0.035cm}i\!\int\frac{d{\bf k}_{1}\, d{\bf k}_{2}\, d{\bf k}_{3}}{(2\pi)^{9}}\,
\biggl\{
\widetilde{T}^{\ \!a\ a_{1}\ a_{2}\ a_{3}}_{{\bf k},\ {\bf k}_{1},\ {\bf k}_{2},\ {\bf k}_{3}}
\biggl(\frac{1}{\Delta\hspace{0.02cm}\omega + i\hspace{0.03cm}0}\biggr)\ \!
\widetilde{T}^{\,*\ \!b\ a_{1}\ a_{2}\ a_{3}}_{{\bf k}^{\prime},\ {\bf k}_{1},\ {\bf k}_{2},\ {\bf k}_{3}}
\]
\[
\times\ \!
(2\pi)^{3}\hspace{0.03cm} \delta({\bf k}^{\prime} + {\bf k}_{1} - {\bf k}_{2} - {\bf k}_{3})\ \!
(2\pi)^{3}\hspace{0.03cm} \delta({\bf k} + {\bf k}_{1} - {\bf k}_{2} - {\bf k}_{3})
\Bigl[\ \!N^{l}_{\bf k\,} N^{l}_{{\bf k}_{2}} N^{l}_{{\bf k}_{3}} +\  \ldots\ \Bigr]
\]
\[
-\
\widetilde{T}^{\ \!a_{2}\ a_{3}\ b\ a_{1}}_{{\bf k_{2}},\ {\bf k}_{3},\ {\bf k}^{\prime},\ {\bf k}_{1}}
\biggl(\frac{1}{\Delta\hspace{0.02cm}\omega - i\hspace{0.03cm}0}\biggr)\ \!
\widetilde{T}^{\,*\ \!a_{2}\ a_{3}\ a\ a_{1}}_{{\bf k_{2}},\ {\bf k}_{3},\ {\bf k},\ {\bf k}_{1}}
\]
\[
\times\ \!
(2\pi)^{3}\hspace{0.03cm} \delta({\bf k} + {\bf k}_{1} - {\bf k}_{2} - {\bf k}_{3})\ \!
(2\pi)^{3}\hspace{0.03cm} \delta({\bf k}^{\prime} + {\bf k}_{1} - {\bf k}_{2} - {\bf k}_{3})
\Bigl[\ \!N^{l}_{\bf k\,} N^{l}_{{\bf k}_{2}} N^{l}_{{\bf k}_{3}} +\  \ldots\ \Bigr]\biggr\}.
\]
Taking into account that
\[
\delta({\bf k}^{\prime} + {\bf k}_{1} - {\bf k}_{2} - {\bf k}_{3})\ \!
\delta({\bf k} + {\bf k}_{1} - {\bf k}_{2} - {\bf k}_{3}) =
\delta({\bf k} - {\bf k}^{\prime})\ \!
\delta({\bf k} + {\bf k}_{1} - {\bf k}_{2} - {\bf k}_{3}),
\]
we can write the last expression in a more compact form:
\[
-\ \! 6\hspace{0.03cm}i (2\pi)^{3}\hspace{0.03cm}\delta({\bf k} - {\bf k}^{\prime})
\!\int\frac{d{\bf k}_{1}\, d{\bf k}_{2}\, d{\bf k}_{3}}{(2\pi)^{9}}\,
(2\pi)^{3}\hspace{0.03cm} \delta({\bf k} + {\bf k}_{1} - {\bf k}_{2} - {\bf k}_{3})
\Bigl[\,N^{l}_{\bf k\,} N^{l}_{{\bf k}_{2}} N^{l}_{{\bf k}_{3}} +\  \ldots\ \Bigr]
\]
\[
\times\ \!
\left\{\ \!
\frac{\widetilde{T}^{\ \!a\ a_{1}\ a_{2}\ a_{3}}_{{\bf k},\ {\bf k}_{1},\ {\bf k}_{2},\ {\bf k}_{3}}\ \!
\widetilde{T}^{\,*\ \!b\ a_{1}\ a_{2}\ a_{3}}_{{\bf k}^{\prime},\ {\bf k}_{1},\ {\bf k}_{2},\ {\bf k}_{3}}}
{\Delta\hspace{0.02cm}\omega + i\hspace{0.03cm}0}
\ -\
\frac{\widetilde{T}^{\ \!a_{2}\ a_{3}\ b\ a_{1}}_{{\bf k_{2}},\ {\bf k}_{3},\ {\bf k}^{\prime},\ {\bf k}_{1}}\ 
\!\widetilde{T}^{\,*\ \!a_{2}\ a_{3}\ a\ a_{1}}_{{\bf k_{2}},\ {\bf k}_{3},\ {\bf k},\ {\bf k}_{1}}}
{\Delta\hspace{0.02cm}\omega - i\hspace{0.03cm}0} \right\}.
\eqno{(4.2)}
\]
Further, performing the convolution of obtained expressions (3.4), (4.1), and (4.2) with $\delta^{a b}$, considering that
\[
\frac{1}{\Delta\hspace{0.02cm}\omega + i\hspace{0.03cm}0}
\ -\,
\frac{1}{\Delta\hspace{0.02cm}\omega - i\hspace{0.03cm}0} = 
-2\hspace{0.02cm}i\hspace{0.03cm}\pi\hspace{0.03cm}\delta(\Delta\hspace{0.02cm}\omega)
\]
and cancelling out the factor $(2\pi)^{3}\hspace{0.03cm}\delta({\bf k} - {\bf k}^{\prime})$, we obtain the desired kinetic equation for colorless longitudinal gluon
\[
\frac{d N^{l}_{{\bf k}}}{d\hspace{0.03cm} t}\
=
\frac{4}{d_{A}}\ \!
N^{l}_{\bf k}\!\int\!\frac{d\hspace{0.03cm}{\bf k}_{1}}{(2\pi)^{3}}\ N^{l}_{{\bf k}_{1}}\ \!
{\rm Im}
\Bigl[\widetilde{T}^{\ a\ a_{1}\ a\ a_{1}}_{{\bf k},\ {\bf k}_{1},\ {\bf k},\ {\bf k}_{1}}\Bigr]
\eqno{(4.3)}
\]
\[
+\; \frac{6}{d_{A}}
\!\int\frac{d\hspace{0.03cm}{\bf k}_{1}\, d\hspace{0.03cm}{\bf k}_{2}\, d\hspace{0.03cm}
{\bf k}_{3}}{(2\pi)^{9}}\, (2\pi)^{4}\hspace{0.03cm}
\delta(\omega^{l}_{{\bf k}} + \omega^{l}_{{\bf k}_{1}} - \omega^{l}_{{\bf k}_{2}} -
\omega^{l}_{{\bf k}_{3}})\ \!
\delta({\bf k} + {\bf k}_{1} - {\bf k}_{2} - {\bf k}_{3})
\]
\[
\times\ \!
\widetilde{T}^{\ \!a\ a_{1}\ a_{2}\ a_{3}}_{{\bf k},\ {\bf k}_{1},\ {\bf k}_{2},\ {\bf k}_{3}}\ \!
\widetilde{T}^{\,*\ \!a\ a_{1}\ a_{2}\ a_{3}}_{{\bf k},\ {\bf k}_{1},\ {\bf k}_{2},\ {\bf k}_{3}}
\Bigl(
N^{l}_{\bf k} N^{l}_{{\bf k}_{2}} N^{l}_{{\bf k}_{3}} +
N^{l}_{{\bf k}_{1}} N^{l}_{{\bf k}_{2}} N^{l}_{{\bf k}_{3}} -
N^{l}_{{\bf k}} N^{l}_{{\bf k}_{1}} N^{l}_{{\bf k}_{3}} -
N^{l}_{{\bf k}} N^{l}_{{\bf k}_{1}} N^{l}_{{\bf k}_{2}}\Bigr).
\]
Here, $d_{A} = N^{2}_{c} - 1$. The first term on the right-hand side of Eq.\,(4.3) describes the so-called nonlinear Landau damping \cite{akhiezer_1974}, the decrement of which is a linear functional of plasmon number density $N^{l}_{\bf k}$:
\[
\hat{\gamma}\bigl\{N^{l}_{\bf k}\bigr\} \equiv \gamma^{l}({\bf k})
=
\frac{4}{d_{A}}\int\!\frac{d\hspace{0.03cm}{\bf k}_{1}}{(2\pi)^{3}}\ N^{l}_{{\bf k}_{1}}\ \!
{\rm Im} \Bigl[\widetilde{T}^{\ a\ a_{1}\ a\ a_{1}}_{{\bf k},\ {\bf k}_{1},\ {\bf k},\ {\bf k}_{1}}\Bigr].
\]
The second term in Eq.\,(4.3) is associated with an elastic plasmon–plasmon scattering. We can also write Eq.\,(4.3) in a more visual form:
\[
\frac{d N_{\bf k}^{l}}{d t} \equiv \frac{\partial N_{\bf k}^{l}}{\partial t} +
{\bf v}_{\bf k}^{l}\cdot\frac{\partial N_{\bf k}^{l}}
{\partial {\bf x}}
=
-\,\hat{\gamma}\,\{ N_{\bf k}^{l} \} \, N_{\bf k}^{l}
- N_{\bf k}^l \Gamma_{\rm d} [ N_{\bf k}^l ] + ( 1 + N_{\bf k}^l )
\Gamma_{\rm i}[ N_{\bf k}^l ]\ ,
\eqno{(4.4)}
\]
where
\[
{\bf v}_{\bf k}^{l} = \frac{\partial \omega_{\bf k}^{l}}
{\partial {\bf k}} = - \Biggl[\left(\frac{\partial {\rm Re} \,
\varepsilon^{l}(k)}{\partial {\bf k}} \right)\!
\left( \frac{\partial {\rm Re}\,\varepsilon^{l}(k)}
{\partial \omega} \right)^{\!-1}\Biggr]\Bigg\vert_{\omega = \omega_{\bf k}^{l}}
\]
is the group velocity of longitudinal oscillations, and the generalized decay rate $\Gamma_{\rm d}$ and the inverse regeneration rate $\Gamma_{\rm i}$ are nonlinear functionals of the plasmon number density:
\[
\Gamma_{\rm d}[N_{\bf k}^l] =
\int\!d{\cal T}^{(3)}{\it w}_{4}({\bf k}, {\bf k}_{1}; {\bf k}_{2}, {\bf k}_{3})
N_{{\bf k}_1}^l (1 + N_{{\bf k}_{2}}^l) (1 + N_{{\bf k}_{3}}^l)
\]
and, accordingly,
\[
\Gamma_{\rm i} [N_{\bf k}^l] =
\int\!d{\cal T}^{(3)}{\it w}_{4}({\bf k}, {\bf k}_{1}; {\bf k}_{2}, {\bf k}_{3})
(1 +  N_{{\bf k}_1}^l) N_{{\bf k}_{2}}^l N_{{\bf k}_{3}}^l\,.
\hspace{0.9cm}
\]
Here,
\[
{\it w}_{4}({\bf k}, {\bf k}_{1}; {\bf k}_{2},{\bf k}_{3}) =
\frac{6}{d_{A}}\
\widetilde{T}^{\ \!a\ a_{1}\ a_{2}\ a_{3}}_{{\bf k},\ {\bf k}_{1},\ {\bf k}_{2},\ {\bf k}_{3}}\ \!
\widetilde{T}^{\,*\, a\ a_{1}\ a_{2}\ a_{3}}_{{\bf k},\ {\bf k}_{1},\ {\bf k}_{2},\ {\bf k}_{3}}
\eqno{(4.5)}
\]
is the scattering probability for an elastic collision of two colorless plasmons, and the integration measure is defined as
\[
d{\cal T}^{(3)} \equiv 
(2\pi)^{4}\,
\delta(\omega^{l}_{{\bf k}} + \omega^{l}_{{\bf k}_{1}} - \omega^{l}_{{\bf k}_{2}} - \omega^{l}_{{\bf k}_{3}})\ \!
\delta({\bf k} + {\bf k}_{1} - {\bf k}_{2} - {\bf k}_{3})\ \!
\frac{d{\bf k}_{1}\, d{\bf k}_{2}\, d{\bf k}_{3}}{(2\pi)^{9}}\ .
\]
In the limit of large occupation numbers of plasmon states ($N_{{\bf k}}^{l}\gg 1$), the right-hand side of Boltzmann equation (4.4) is transformed into (4.3).


\section{Explicit form of the function $T^{\; a\; a_{1}\; a_{2}\; a_{3}}_{{\bf k},\, {\bf k}_{1},\, {\bf k}_{2},\, {\bf k}_{3}}$}

It remains for us to determine the explicit form of vertex functions $T^{\ a\ a_{1}\ a_{2}\ a_{3}}_{{\bf k},\ {\bf k}_{1},\ {\bf k}_{2},\ {\bf k}_{3}}$,  $U^{\ \! a\ a_{1}\ a_{2}}_{{\bf k},\ {\bf k}_{1},\ {\bf k}_{2}}$, and 
$V^{\ \! a\ a_{1}\ a_{2}}_{{\bf k},\ {\bf k}_{1},\ {\bf k}_{2}}$ that appear in effective amplitude (2.13). In this section, we determine the form of function $T^{\ a\ a_{1}\ a_{2}\ a_{3}}_{{\bf k},\ {\bf k}_{1},\ {\bf k}_{2},\ {\bf k}_{3}}$ in the hard
thermal loop (HTL) approximation \cite{blaizot_2002}. In \cite{markov_2002}, the probability of elastic scattering of two plasmons was determined in the HTL approximation:
\[
{\it w}_{4}({\bf k}, {\bf k}_{1}; {\bf k}_{2},{\bf k}_{3})\! =\!3\hspace{0.03cm}
{\rm M}^{\,a\, a_{1}\, a_{2}\, a_{3}}({\bf k}, {\bf k}_1,-{\bf k}_2,-{\bf k}_3)\hspace{0.03cm}
{\rm M}^{\hspace{0.02cm}*\ \!a\, a_{1}\, a_{2}\, a_{3}}({\bf k}, {\bf k}_1,-{\bf k}_2,-{\bf k}_3).
\eqno{(5.1)}
\]
Here, the matrix element of the four-plasmon decay has the following structure:
\[
{\rm M}^{\,a\, a_{1}\, a_{2}\, a_{3}}({\bf k}, {\bf k}_1,-{\bf k}_2,-{\bf k}_3)
= g^{2}\!\left(\frac{{\rm Z}_l({\bf k})}{2\omega_{\bf k}^l}\right)^{\!1/2}\!
\!\left(\frac{\tilde{u}^{\mu}(k)}{\sqrt{\bar{u}^2(k)}}\right)
\hspace{1cm}
\eqno{(5.2)}
\]
\[
\hspace{1cm}
\times\,\prod_{i=1}^{3}
\left(\frac{{\rm Z}_l({\bf k}_i)}{2\omega_{{\bf k}_i}^l}\right)^{\!1/2}\!
\!\left(\frac{\tilde{u}^{\mu_i}(k_i)}{\sqrt{\bar{u}^2(k_i)}}\right)
\,^{\ast}\widetilde{\Gamma}^{\,a\, a_{1}\, a_{2}\, a_{3}}_{\,\mu\, \mu_{1}\, \mu_{2}\, \mu_{3}}
(k,k_{1},-k_2, -k_3)\Big|_{\rm on-shell}
\]
and, in turn, the effective amplitude $\!\,^{\ast}\widetilde{\Gamma}^{\,a\, a_{1}\, a_{2}\, a_{3}}_{\,\mu\, \mu_{1}\, \mu_{2}\, \mu_{3}}(k,k_{1},-k_{2},-k_3)$ is defined as
\[
\,^{\ast}\widetilde{\Gamma}^{\,a\, a_{1}\, a_{2}\, a_{3}}_{\,\mu\, \mu_{1}\, \mu_{2}\, \mu_{3}}(k,k_{1},-k_{2},-k_3) =
-f^{\ \!\!a\ \!\!a_{1}\ \!\!b}f^{\ \!\!b\ \!\!a_{2}\ \!\!a_{3}}
\,^{\ast}\widetilde{\Gamma}_{\,\mu\, \mu_{1}\, \mu_{2}\, \mu_{3}}(k,k_{1},-k_{2},-k_3)
\eqno{(5.3)}
\]
\[
-\,f^{\ \!\!a\ \!\!a_{2}\ \!\!b}f^{\ \!\!b\ \!\!a_{1}\ \!\!a_{3}}
\,^{\ast}\widetilde{\Gamma}_{\,\mu\, \mu_{2}\, \mu_{1}\, \mu_{3}}(k,-k_{2},k_{1},-k_3),
\]
where $f^{\,a\,b\,c}$ are antisymmetric structural constants of the color Lie algebra $\mathfrak{su}(N_c)$. Color factors in this expression are multiplied by the purely kinetic coefficients, viz., effective subamplitudes defined as
\[
\,^{\ast}\widetilde{\Gamma}_{\,\mu\, \mu_{1}\, \mu_{2}\, \mu_{3}}(k,k_{1},-k_{2},-k_3)
\equiv
\,^{\ast}{\Gamma}_{\,\mu\, \mu_{1}\, \mu_{2}\, \mu_{3}}(k,k_{1},-k_{2},-k_3)
\eqno{(5.4)}
\]
\vspace{-0.35cm}
\[
- \,^{\ast}\Gamma_{\,\mu\, \mu_{1}\, \nu}(k,k_1,-k - k_1)
\,^{\ast}\widetilde{\cal D}^{\,\nu\nu^{\prime}}(k_2+k_3)
\,^{\ast}\Gamma_{\,\nu^{\prime}\, \mu_{2}\, \mu_{3}}(k_2+k_3,-k_2,-k_3)
\]
\vspace{-0.35cm}
\[
- \,^{\ast}\Gamma_{\,\mu\,\mu_{3}\, \nu}(k,-k_3,-k+k_3)
\,^{\ast}\widetilde{\cal D}^{\,\nu\nu^{\prime}}(k_{2} - k_{1})
\,^{\ast}\Gamma_{\,\nu^{\prime}\, \mu_{2}\, \mu_{1}}(k_{2} - k_{1},-k_2,k_1).
\hspace{0.4cm}
\]
The form of vertex functions $\!\,^{\ast}{\Gamma}_{\,\mu\, \mu_{1}\, \mu_{2}\, \mu_{3}}(k,k_{1},k_{2},k_3)$ and $\!\,^{\ast}\Gamma_{\,\mu\, \mu_{1}\, \mu_{2}}(k,k_{1},k_{2})$, (A.1)\,--\,(A.7), as well as of gluon propagator $\,^{\ast}\!\tilde{\cal D}^{\nu\mu}(k)$, is given in Appendix in the HTL approximation, (A.8)\,--\,(A.10). Two four-vectors 
\[
\tilde{u}_{\mu} (k) = \frac{k^2}{(k\cdot u)}\ \! \Bigl(k_{\mu} - u_{\mu}(k\cdot u)\Bigr)
\quad \mbox{и} \quad
\bar{u}_{\mu} (k) = k^2 u_{\mu} - k_{\mu}(k\cdot u)
\eqno{(5.5)}
\]
are the projectors onto the longitudinal direction of wavevector ${\bf k}$, written in the Lorentz-invariant form in the Hamilton and Lorentz gauge, respectively. Here, $u^{\mu}$ is the four-velocity of the medium, which is $u^{\mu}=(1,0,0,0)$ in the rest system. 
Finally, four-vectors of form
\[
\left(\frac{{\rm Z}_l({\bf k})}{2\omega_{\bf k}^l}\right)^{\!1/2}\!\!\!
\left.\frac{\tilde{u}_{\mu}(k)}{\sqrt{\bar{u}^2(k)}}\ \!\right|_{\rm on-shell}\!
\equiv\,
\frac{1}{\sqrt{2\omega^l_{\bf k}}}\ \epsilon^{l}_{\mu}({\bf k})
\eqno{(5.6)}
\]
on the right-hand side of Eq.\,(5.2) are conventional wavefunctions of a longitudinal physical gluon in the $A_0$\hspace{0.03cm}-\hspace{0.03cm}gauge, where factor $\sqrt{{\rm Z}_l({\bf k})}$ ensures renormalization of the gluon wavefunction due to thermal effects. Factor 3 on the right-hand side of expression (5.1) accounts for three possible four-plasmon decay channels, which change the plasmon number density:
\[
{\rm g}^{\ast}+{\rm g}_1^{\ast}
\rightleftharpoons{\rm g}_2^{\ast}+{\rm g}_3^{\ast},\quad
{\rm g}^{\ast}+{\rm g}_2^{\ast}
\rightleftharpoons{\rm g}_1^{\ast}+{\rm g}_3^{\ast},\quad
{\rm g}^{\ast}+{\rm g}_3^{\ast}
\rightleftharpoons{\rm g}_1^{\ast}+{\rm g}_2^{\ast}.
\]
\indent Comparing two expressions (4.5) and (5.1) for the plasmon–plasmon scattering probability, we see that effective amplitude $\widetilde{T}^{\,a\ a_{1}\ a_{2}\ a_{3}}_{\,{\bf k},\ {\bf k}_{1},\ {\bf k}_{2},\ {\bf k}_{3}}$ defined by expression (2.13) should be identified (to within a numerical factor) with matrix element 
${\rm M}^{\,a\, a_{1}\, a_{2}\, a_{3}}({\bf k}, {\bf k}_1,-{\bf k}_2,-{\bf k}_3)$, calculated using the high-temperature quantum field theory; i.e.,
\[
\widetilde{T}^{\ a\ a_{1}\ a_{2}\ a_{3}}_{{\bf k},\ {\bf k}_{1},\ {\bf k}_{2},\ {\bf k}_{3}}
=
\biggl(\frac{d_{A}}{2}\biggr)^{\!1/2\!}{\rm M}^{\,a\, a_1\, a_2\, a_3}({\bf k}, {\bf k}_1,-{\bf k}_2,
-{\bf k}_3).
\eqno{(5.7)}
\]
From expressions for effective amplitudes (2.13) and (5.2), (5.3), we can immediately obtain the explicit form of amplitude $T^{\ \!a\ a_{1}\ a_{2}\ a_{3}}_{{\bf k},\ {\bf k}_{1},\ {\bf k}_{2},\ {\bf k}_{3}}$, which appears as the coefficient function in the definition of fourth-order Hamiltonian $\widehat{H}_{4}$ (2.7):
\[
T^{\ \!a\ a_{1}\ a_{2}\ a_{3}}_{{\bf k},\ {\bf k}_{1},\ {\bf k}_{2},\ {\bf k}_{3}}
=
- \biggl(\frac{d_{A}}{2}\biggr)^{\!\!1/2\!}g^{2} \Biggl(
\frac{ \epsilon^{l}_{\mu}({\bf k})}{\sqrt{2\omega^l_{\bf k}}}\Biggr)
\prod_{i=1}^{3}\ \!
\Biggl(\frac{\epsilon^{l}_{\mu_{i}}({\bf k}_{i})}{\sqrt{2\omega^l_{{\bf k}_{i}}}}\Biggr)
\ \!\times
\]
\[
\times
\Bigl[f^{\ \!\!a\ \!\!a\ \!\!_{1}b}f^{\ \!\!b\ \!\!a_{2}\ \!\!a_{3}}
\,^{\ast}{\Gamma}^{\,\mu\, \mu_{1}\, \mu_{2}\, \mu_{3}}(k,k_{1},\!-k_{2},\!-k_3)
+\!
\,f^{\ \!\!a\ \!\!a_{2}\ \!\!b}f^{\ \!\!b\ \!\!a_{1}\ \!\!a_{3}}
\,^{\ast}{\Gamma}^{\,\mu\, \mu_{2}\, \mu_{1}\, \mu_{3}}(k,\!-k_{2},k_{1},\!-k_3)\Bigr]\!
\Big|_{\rm on\!-shell}\!.
\eqno{(5.8)}
\]
Here, we have taken into account the relationship (5.6) of the longitudinal projector with the polarization vector. The explicit form of effective four-gluon vertex $\,^{\ast} \Gamma^{\hspace{0.02cm}\mu\hspace{0.02cm} \nu\hspace{0.02cm} \lambda\hspace{0.02cm} \sigma}(k,k_1,k_2,k_3)$ on the right-hand side of expression (5.8) is defined by formulas (A.5)\,--\,(A.7).


\section{Explicit form of functions $U^{\ \! a\, a_{1}\, a_{2}}_{{\bf k},\, {\bf k}_{1},\, {\bf k}_{2}}$ and
$V^{\ \! a\, a_{1}\, a_{2}}_{{\bf k},\, {\bf k}_{1},\, {\bf k}_{2}}$}
\setcounter{equation}{0}

Let us now determine the explicit form of coefficient functions   $U^{\ \! a\ a_{1}\ a_{2}}_{{\bf k},\ {\bf k}_{1},\ {\bf k}_{2}}$ and $V^{\ \! a\ a_{1}\ a_{2}}_{{\bf k},\ {\bf k}_{1},\ {\bf k}_{2}}$ in the integrands of third-order Hamiltonian $\widehat{H}_{3}$ (2.6). In contrast to the previous case, however, here we have a more complicated situation. Considering relations (2.13) and (5.2)\,--\,(5.4), we obtain from formula (5.7) the following initial expression for analysis:
\begin{align}
&\frac{U^{\ b\ a_{2}\ a_{3}}_{-({\bf k}_{2}+{\bf k}_{3}),\ {\bf k}_{2},\ {\bf k}_{3}}\
U^{*\ b\ a\ a_{1}}_{-({\bf k}+{\bf k}_{1}),\ {\bf k},\ {\bf k}_{1}}}
{\omega^{l}_{-({\bf k}+{\bf k}_{1})\!} + \omega^{l}_{{\bf k}} + \omega^{l}_{{\bf k}_{1}}}
\ +\ \frac{V^{\ b\ a_{2}\ a_{3}}_{{\bf k}_{2}+{\bf k}_{3},\ {\bf k}_{2},\ {\bf k}_{3}}\
V^{*\ b\ a\ a_{1}}_{{\bf k} + {\bf k}_{1},\ {\bf k},\ {\bf k}_{1}}}
{\omega^{l}_{{\bf k}+{\bf k}_{1}\!} - \omega^{l}_{{\bf k}} - \omega^{l}_{{\bf k}_{1}}}
\notag\\[1ex]
&+\ \!\frac{V^{\ a_{1}\ a_{2}\ b}_{{\bf k}_{1},\ {\bf k}_{2},\
{\bf k}_{1}-{\bf k}_{2}}\ V^{*\ a_{3}\ a\ b}_{{\bf k}_{3},\ {\bf k},\ {\bf k}_{3} - {\bf k}}}
{\omega^{l}_{{\bf k}_{3}-{\bf k}\!} + \omega^{l}_{{\bf k}} - \omega^{l}_{{\bf k}_{3}}}
\ +\ \frac{V^{\ a\ a_{2}\ b}_{{\bf k},\ {\bf k}_{2},\ {\bf k}-{\bf k}_{2}}\
V^{*\ a_{3}\ a_{1}\ b}_{{\bf k}_{3},\ {\bf k}_{1},\ {\bf k}_{3}-{\bf k}_{1}}}
{\omega^{l}_{{\bf k}_{3}-{\bf k}_{1}\!} + \omega^{l}_{{\bf k}_{1}} - \omega^{l}_{{\bf k}_{3}}}
\notag\\[1ex]
&+\ \!\frac{V^{\ a\ a_{3}\ b}_{{\bf k},\ {\bf k}_{3},\ {\bf k}-{\bf k}_{3}}\
V^{*\ a_{2}\ a_{1}\ b}_{{\bf k}_{2},\ {\bf k}_{1},\ {\bf k}_{2}-{\bf k}_{1}}}
{\omega^{l}_{{\bf k}_{2} - {\bf k}_{1}\!} + \omega^{l}_{{\bf k}_{1}} - \omega^{l}_{{\bf k}_{2}}}
\ +\
\frac{V^{\ a_{1}\ a_{3}\ b}_{{\bf k}_{1},\ {\bf k}_{3},\ {\bf k}_{1}-{\bf k}_{3}}\
V^{*\ a_{2}\ a\ b}_{{\bf k}_{2},\ {\bf k},\ {\bf k}_{2}-{\bf k}}}
{\omega^{l}_{{\bf k}_{2}-{\bf k}\!} + \omega^{l}_{{\bf k}} - \omega^{l}_{{\bf k}_{2}}}
\notag
\end{align}
\[
=
\frac{1}{2}\, \biggl(\frac{d_{A}}{2}\biggr)^{\!\!1/2\!}g^{2} \Biggl(
\frac{ \epsilon^{l}_{\mu}(k)}{\sqrt{2\omega^l_{\bf k}}}\Biggr)
\prod_{i=1}^{3}\ \!
\Biggl(\frac{\epsilon^{l}_{\mu_{i}}(k_{i})}{\sqrt{2\omega^l_{{\bf k}_{i}}}}\Biggr)
\ \!\times
\eqno{(6.1)}
\]
\[
\Bigl[\,f^{\ \!\!a\ \!\!a_{1}\ \!\!b}f^{\ \!\!b\ \!\!a_{2}\ \!\!a_{3}}
\Bigl(\!\,^{\ast}\Gamma^{\mu\mu_1\nu}(k,k_1,-k - k_1)
\,^{\ast}\widetilde{\cal D}_{\nu\nu^{\prime}}(k_2+k_3)
\,^{\ast}\Gamma^{\nu^{\prime}\mu_2\mu_3}(k_2+k_3,-k_2,-k_3)
\]
\[
+\, \,^{\ast}\Gamma^{\mu\mu_3\nu}(k,-k_3,-k+k_3)
\,^{\ast}\widetilde{\cal D}_{\nu\nu^{\prime}}(k_{2} - k_{1})
\,^{\ast}\Gamma^{\nu^{\prime}\mu_2\mu_1}(k_{2} - k_{1},-k_2,k_1)\Bigr)
\]
\[
+\,f^{\ \!\!a\ \!\!a_{2}\ \!\!{b}}f^{\ \!\!b\ \!\!a_{1}\ \!\!a_{3}}
\Bigl(\!\,^{\ast}\Gamma^{\mu\mu_1\nu}(k,-k_2,-k + k_2)
\,^{\ast}\widetilde{\cal D}_{\nu\nu^{\prime}}(- k_1+k_3)
\,^{\ast}\Gamma^{\nu^{\prime}\mu_2\mu_3}(-k_1+k_3,k_1,-k_3)
\]
\[
+ \;^{\ast}\Gamma^{\mu\mu_3\nu}(k,-k_3,-k+k_3)
\,^{\ast}\widetilde{\cal D}_{\nu\nu^{\prime}}(-k_{1} + k_{2})
\,^{\ast}\Gamma^{\nu^{\prime}\mu_2\mu_1}(-k_{1} + k_{2},k_1, - k_2)\Bigr)
\Bigr]\Big|_{\rm on-shell}.
\]
At the first step, we must ``untangle'' the color structure of this expression. For this, we set for three-point amplitudes $U$ and $V$:
\[
U^{\ \! a\ a_{1}\ a_{2}}_{{\bf k},\ {\bf k}_{1},\ {\bf k}_{2}}
=
f^{\ \!\!a\ \!\!a_{1}\ \!\!a_{2}\,}U_{{\bf k},\ {\bf k}_{1},\ {\bf k}_{2}},
\qquad
V^{\ \! a\ a_{1}\ a_{2}}_{{\bf k},\ {\bf k}_{1},\ {\bf k}_{2}}
=
f^{\ \!\!a\ \!\!a_{1}\ \!\!a_{2}\,}V_{{\bf k},\ {\bf k}_{1},\ {\bf k}_{2}}.
\]
Such a representation is unambiguous. In view of complete antisymmetry of the structure constants $f^{\ \!\!a\ \!\! a_{1}\ \!\! a_{2}}$ in permutation of color indices, properties (2.8) immediately imply
$$
V_{{\bf k},\ {\bf k}_{1},\ {\bf k}_{2}} = -\ \! V_{{\bf k},\ {\bf k}_{2},\ {\bf k}_{1}}\ ,
\quad U_{{\bf k},\ {\bf k}_{1},\ {\bf k}_{2}} = -\ \! U_{{\bf k},\ {\bf k}_{2},\ {\bf k}_{1}}
= U_{{\bf k}_{1},\ {\bf k}_{2},\ {\bf k}}\, .
\eqno{(6.2)}
$$
\indent Further, using the identity
\[
f^{\ \!\!a_{1}\ \!\!a_{2}\ \!\!b}f^{\ \!\!b\ \!\!a_{3}\ \!\!a} =
- f^{\ \!\!a\ \!\!a_{2}\ \!\!b}f^{\ \!\!b\ \!\!a_{1}\ \!\!a_{3}} + f^{\ \!\!a\ \!\!a_{1}\ \!\!b}f^{\ \!\!b\ \!\!a_{2}\ \!\!a_{3}},
\]
for antisymmetric structure constants, we can reduce the left-hand side of relation (6.1) to form
\[
f^{\ \!\!a\ \!\!a_{1}\ \!\!b}f^{\ \!\!b\ \!\!a_{2}\ \!\!a_{3}}\Biggl[\
\frac{U^{\phantom{*}}_{-({\bf k}_{2}+{\bf k}_{3}),\ {\bf k}_{2},\ {\bf k}_{3}}\
U^{*}_{-({\bf k}+{\bf k}_{1}),\ {\bf k},\ {\bf k}_{1}}}
{\omega^{l}_{-({\bf k}+{\bf k}_{1})} + \omega^{l}_{{\bf k}} + \omega^{l}_{{\bf k}_{1}}}
\ +\
\frac{V^{\phantom{*}}_{{\bf k}_{2}+{\bf k}_{3},\ {\bf k}_{2},\ {\bf k}_{3}}\
V^{*}_{{\bf k} + {\bf k}_{1},\ {\bf k},\ {\bf k}_{1}}}
{\omega^{l}_{{\bf k} + {\bf k}^{l}_{1}} - \omega^{l}_{{\bf k}} - \omega^{l}_{{\bf k}_{1}}}
\]
\[
\hspace{4.5cm}
+\ \!\frac{V^{\phantom{*}}_{{\bf k}_{1},\ {\bf k}_{2},\ {\bf k}_{1}-{\bf k}_{2}}\ V^{*}_{{\bf k}_{3},\
{\bf k},\ {\bf k}_{3}-{\bf k}}
}{\omega_{{\bf k}_{3}-{\bf k}}+\omega_{{\bf k}}-\omega_{{\bf k}_{3}}}
\ +\
\!\frac{V^{\phantom{*}}_{{\bf k},\ {\bf k}_{3},\ {\bf k}-{\bf k}_{3}}\ V^{*}_{{\bf k}_{2},\
{\bf k}_{1},\ {\bf k}_{2}-{\bf k}_{1}}
}{\omega_{{\bf k}_{2}-{\bf k}_{1}}+\omega_{{\bf k}_{1}}-\omega_{{\bf k}_{2}}}\ \Biggr] 
\]
\[
- f^{\ \!\!a\ \!\!a_{2}\ \!\!b}f^{\ \!\!b\ \!\!a_{1}\ \!\!a_{3}}\Biggl[\
\!\frac{V^{\phantom{*}}_{{\bf k}_{1},\ {\bf k}_{2},\ {\bf k}_{1}-{\bf k}_{2}}\ V^{*}_{{\bf k}_{3},\
{\bf k},\ {\bf k}_{3}-{\bf k}}
}{\omega_{{\bf k}_{3}-{\bf k}}+\omega_{{\bf k}}-\omega_{{\bf k}_{3}}}
\ +\ \frac{V^{\phantom{*}}_{{\bf k},\ {\bf k}_{2},\ {\bf k}-{\bf k}_{2}}\ V^{*}_{{\bf k}_{3},\
{\bf k}_{1},\ {\bf k}_{3}-{\bf k}_{1}}
}{\omega_{{\bf k}_{3}-{\bf k}_{1}}+\omega_{{\bf k}_{1}}-\omega_{{\bf k}_{3}}}
\hspace{1.4cm}
\]
\[
\hspace{4.5cm}
+\
\frac{V^{\phantom{*}}_{{\bf k},\ {\bf k}_{3},\ {\bf k}-{\bf k}_{3}}\ V^{*}_{{\bf k}_{2},\
{\bf k}_{1},\ {\bf k}_{2}-{\bf k}_{1}}
}{\omega_{{\bf k}_{2}-{\bf k}_{1}}+\omega_{{\bf k}_{1}}-\omega_{{\bf k}_{2}}}
\ +\
\frac{V^{\phantom{*}}_{{\bf k}_{1},\ {\bf k}_{3},\ {\bf k}_{1}-{\bf k}_{3}}\ V^{*}_{{\bf k}_{2},\
{\bf k},\ {\bf k}_{2}-{\bf k}}
}{\omega_{{\bf k}_{2}-{\bf k}}+\omega_{{\bf k}}-\omega_{{\bf k}_{2}}}\ \Biggr].
\]
Comparing the coefficients following products $f^{\ \!\!a\ \!\!a_{1}\ \!\!b}f^{\ \!\!b\ \!\!a_{2}\ \!\!a_{3}}$ and $ f^{\ \!\!a\ \!\!a_{2}\ \!\!b}f^{\ \!\!b\ \!\!a_{1}\ \!\!a_{3}}$  in the above expression and the right-hand side of relation (6.1), we obtain
\[
\frac{U^{\phantom{*}}_{-({\bf k}_{2}+{\bf k}_{3}),\ {\bf k}_{2},\ {\bf k}_{3}}\
U^{*}_{-({\bf k}+{\bf k}_{1}),\ {\bf k},\ {\bf k}_{1}}}{\omega_{-({\bf k}+{\bf k}_{1})}
+ \omega_{{\bf k}}+\omega_{{\bf k}_{1}}}
\ +\
\frac{V^{\phantom{*}}_{{\bf k}_{2}+{\bf k}_{3},\ {\bf k}_{2},\ {\bf k}_{3}}\ V^{*}_{{\bf k}
+
{\bf k}_{1},\ {\bf k},\ {\bf k}_{1}}}{\omega_{{\bf k}+{\bf k}_{1}}-\omega_{{\bf k}}-
\omega_{{\bf k}_{1}}}
\]
\[
\hspace{0.5cm}
+\ \frac{V^{\phantom{*}}_{{\bf k}_{1},\ {\bf k}_{2},\ {\bf k}_{1}-{\bf k}_{2}}\
V^{*}_{{\bf k}_{3},\ {\bf k},\ {\bf k}_{3}-{\bf k}}}
{\omega_{-({\bf k}_{2}-{\bf k}_{1})} + \omega_{{\bf k}_{2}}-\omega_{{\bf k}_{1}}}
\ +\
\!\frac{V^{\phantom{*}}_{{\bf k},\ {\bf k}_{3},\ {\bf k}-{\bf k}_{3}}\ V^{*}_{{\bf k}_{2},\
{\bf k}_{1},\ {\bf k}_{2}-{\bf k}_{1}}
}{\omega_{{\bf k}_{2}-{\bf k}_{1}}+\omega_{{\bf k}_{1}}-\omega_{{\bf k}_{2}}}
\hspace{1.2cm}
\]
\[
=
\frac{1}{2}\, \biggl(\frac{d_{A}}{2}\biggr)^{\!\!1/2\!}g^{2} \Biggl(
\frac{ \epsilon^{l}_{\mu}({\bf k})}{\sqrt{2\omega^l_{\bf k}}}\Biggr)
\prod_{i=1}^{3}\ \!
\Biggl(\frac{\epsilon^{l}_{\mu_{i}}({\bf k}_{i})}{\sqrt{2\omega^l_{{\bf k}_{i}}}}\Biggr)
\ \!\times
\]
\[
\times
\Bigl[\,^{\ast}\Gamma^{\mu\mu_1\nu}(k,k_1,-k - k_1)
\,^{\ast}\widetilde{\cal D}_{\nu\nu^{\prime}}(k + k_1)
\,^{\ast}\Gamma^{\nu^{\prime}\mu_2\mu_3}(k_2+k_3,-k_2,-k_3)
\hspace{1.5cm}
\]
\[
+\, \,^{\ast}\Gamma^{\mu\mu_3\nu}(k,-k_3,-k+k_3)
\,^{\ast}\widetilde{\cal D}_{\nu\nu^{\prime}}(k_{2} - k_{1})
\,^{\ast}\Gamma^{\nu^{\prime}\mu_2\mu_1}(k_{2} - k_{1},-k_2,k_1)\Bigr]\Bigr|_{\rm \,on-shell}
\]
plus the analogous relation for the second coefficient function. Finally, the structure of this relation clearly shows that in fact we have here two independent relations: first,
\[
\frac{U^{\phantom{*}}_{-({\bf k}_{2}+{\bf k}_{3}),\ {\bf k}_{2},\ {\bf k}_{3}}\
U^{*}_{-({\bf k}+{\bf k}_{1}),\ {\bf k},\ {\bf k}_{1}}}
{\omega^l_{{\bf k}+{\bf k}_{1}} + \omega^l_{{\bf k}} + \omega^l_{{\bf k}_{1}}}
\ +\
\frac{V^{\phantom{*}}_{{\bf k}_{2}+{\bf k}_{3},\ {\bf k}_{2},\ {\bf k}_{3}}\ V^{*}_{{\bf k}
+
{\bf k}_{1},\ {\bf k},\ {\bf k}_{1}}}{\omega^l_{{\bf k}+{\bf k}_{1}}-\omega^l_{{\bf k}}-
\omega^l_{{\bf k}_{1}}}
\eqno{(6.3)}
\]
\[
=
\frac{1}{2}\, \biggl(\frac{d_{A}}{2}\biggr)^{\!\!1/2\!}g^{2} \Biggl(
\frac{ \epsilon^{l}_{\mu}({\bf k})}{\sqrt{2\omega^l_{\bf k}}}\Biggr)
\prod_{i=1}^{3}\ \!
\Biggl(\frac{\epsilon^{l}_{\mu_{i}}({\bf k}_{i})}{\sqrt{2\omega^l_{{\bf k}_{i}}}}\Biggr)
\]
\[
\times
\Bigl[\,^{\ast}\Gamma^{\mu\mu_1\nu}(k,k_1,-k - k_1)
\,^{\ast}\widetilde{\cal D}_{\nu\nu^{\prime}}(k + k_1)
\,^{\ast}\Gamma^{\nu^{\prime}\mu_2\mu_3}(k_2+k_3,-k_2,-k_3)\Bigr]\Bigr|_{\rm \,on-shell}
\]
and second,
\[
\frac{V^{\phantom{*}}_{{\bf k}_{1},\ {\bf k}_{2},\ {\bf k}_{1}-{\bf k}_{2}}\
V^{*}_{{\bf k}_{3},\ {\bf k},\ {\bf k}_{3}-{\bf k}}}
{\omega^l_{{\bf k}_{2}-{\bf k}_{1}} + \omega^l_{{\bf k}_{2}} - \omega^l_{{\bf k}_{1}}}
\ +\
\!\frac{V^{\phantom{*}}_{{\bf k},\ {\bf k}_{3},\ {\bf k}-{\bf k}_{3}}\ V^{*}_{{\bf k}_{2},\
{\bf k}_{1},\ {\bf k}_{2}-{\bf k}_{1}}}
{\omega^l_{{\bf k}_{2}-{\bf k}_{1}}+\omega^l_{{\bf k}_{1}} - \omega^l_{{\bf k}_{2}}}\  =
\hspace{1.2cm}
\eqno{(6.4)}
\]
\[
=
\frac{1}{2}\, \biggl(\frac{d_{A}}{2}\biggr)^{\!\!1/2\!}g^{2} \Biggl(
\frac{ \epsilon^{l}_{\mu}({\bf k})}{\sqrt{2\omega^l_{\bf k}}}\Biggr)
\prod_{i=1}^{3}\ \!
\Biggl(\frac{\epsilon^{l}_{\mu_{i}}({\bf k}_{i})}{\sqrt{2\omega^l_{{\bf k}_{i}}}}\Biggr)
\]
\[
\times
\Bigl[\,^{\ast}\Gamma^{\mu\mu_3\nu}(k,-k_3,-k+k_3)
\,^{\ast}\widetilde{\cal D}_{\nu\nu^{\prime}}(k_{2} - k_{1})
\,^{\ast}\Gamma^{\nu^{\prime}\mu_2\mu_1}(k_{2} - k_{1},-k_2,k_1)\Bigr]\Bigr|_{\rm \,on-shell}.
\]
On the left-hand sides of expressions (6.3) and (6.4), we have taken into account the evenness of the dispersion relation (i.e.,
$\omega^l_{-{\bf k}} =  \omega^l_{{\bf k}}$).\\
\indent Further, at the second step in effective gluon propagators $\,^{\ast}\widetilde{\cal D}_{\nu\nu^{\prime}}$, on the right-hand sides of relations (6.3) and (6.4) we retain only the terms with longitudinal projector $\widetilde{Q}_{\nu\nu^{\prime}}$. For example, for first propagator $\,^{\ast}\widetilde{\cal D}_{\nu\nu^{\prime}} (k + k_1)$, we perform substitution
\[
\,^{\ast}\!\widetilde{\cal D}_{\nu\nu^{\prime}}(k + k_1)
\Rightarrow
-\, \widetilde{Q}_{\nu\nu^{\prime}}(k + k_1) \,^{\ast}\!\Delta^l(k + k_1),
\]
where the right-hand side, on account of relations (A.10) and (A.9), is given by explicit expression
\[
-\,\frac{\tilde{u}_{\nu}(k + k_1)\, \tilde{u}_{\nu^{\prime}}(k + k_1)}
{\bar{u}^2(k + k_1)}\ \frac{1}{(k + k_1)^2 - \Pi^{l}(k + k_1)},
\eqno{(6.5)}
\]
analogous operations are performed for second propagator $\,^{\ast}\widetilde{\cal D}_{\nu\nu^{\prime}}(k_{2} - k_{1})$. Near pole $\omega\sim\omega^{l}_{\bf k}$, longitudinal scalar propagator $\,^{\ast}\!\Delta^l(k) = \,^{\ast}\!\Delta^l(\omega,\,{\bf k})$ behaves as (see, for example, \cite{weldon_1998})
\[
\,^{\ast}\!\Delta^l(\omega,\,{\bf k}) = \frac{1}{\omega^2 - {\bf k}^2 - \Pi^{l}(\omega,\,{\bf k})}
\simeq
\frac{{\rm Z}_l({\bf k})}{\omega^{2} - (\omega^{l}_{\bf k})^{2}}
=
\biggl(\frac{{\rm Z}_l({\bf k})}{2\hspace{0.03cm}\omega^{l}_{\bf k}}\biggr)
\biggl[\,\frac{1}{\omega - \omega^{l}_{\bf k}} - \frac{1}{\omega + \omega^{l}_{\bf k}}\,\biggr] .
\]
Using this approximation, we obtain, in particular, the following expressions for the first propagator:
\[
^{\ast}\!\Delta^l(k + k_1)
\simeq
-\,\biggl(\frac{{\rm Z}_l({\bf k} + {\bf k}_{1})}{2\ \!\omega^{l}_{{\bf k} + {\bf k}_{1}}}\biggr)
\biggl[\,\frac{1}{\omega^{l}_{{\bf k} + {\bf k}_{1}}\! - \omega^{l}_{{\bf k}} -
\omega^{l}_{{\bf k}_{1}}}
+
\frac{1}{\omega^{l}_{{\bf k} + {\bf k}_{1}}\! + \omega^{l}_{{\bf k}} +
\omega^{l}_{{\bf k}_{1}}}\,\biggr]
\hspace{0.35cm}
\eqno{(6.6)}
\]
and for the second propagator
\[
^{\ast}\!\Delta^l(k_2 - k_1)
\simeq
-\,\biggl(\frac{{\rm Z}_l({\bf k}_{2} - {\bf k}_{1})}{2\ \!\omega^{l}_{{\bf k}_{2} - {\bf k}_{1}}}\biggr) \biggl[\,
\frac{1}{\omega^{l}_{{\bf k}_{2} - {\bf k}_{1}}\! + \omega^{l}_{{\bf k}_{2}} -
\omega^{l}_{{\bf k}_{1}}}
+
\frac{1}{\omega^{l}_{{\bf k}_{2} - {\bf k}_{1}}\! + \omega^{l}_{{\bf k}_{1}} -
\omega^{l}_{{\bf k}_{2}}}\,\biggr].
\eqno{(6.7)}
\]
Considering expressions (6.3)\,--\,(6.7) given above, we can write the desired three-plasmon vertex functions in the form 
\[
V_{{\bf k},\ {\bf k}_{1},\ {\bf k}_{2}} =
g\,\biggl(\frac{d_{A}}{8}\biggr)^{\!\!1/4\!}
\Biggl(\frac{ \epsilon^{l}_{\mu}({\bf k})}{\sqrt{2\omega^l_{{\bf k}_{\phantom{1}}}}}\Biggr)\!
\Biggl(\frac{\epsilon^{l}_{\mu_{1}}({\bf k}_{1})}{\sqrt{2\omega^l_{{\bf k}_{1}}}}\Biggr)\!
\Biggl(\frac{\epsilon^{l}_{\mu_{2}}({\bf k}_{2})}{\sqrt{2\omega^l_{{\bf k}_{2}}}}\Biggr)\!
\,^{\ast}\Gamma^{\mu\mu_1\mu_2}(k,- k_{1},- k_{2})\Bigr|_{\rm \,on-shell}
\hspace{0.3cm}
\eqno{(6.8)}
\]
and
\[
U_{{\bf k},\ \!{\bf k}_{1},\ \!{\bf k}_{2}} =
g\,\biggl(\frac{d_{A}}{8}\biggr)^{\!\!1/4\!}
\Biggl(\frac{ \epsilon^{l}_{\mu}({\bf k})}{\sqrt{2\omega^l_{{\bf k}_{\phantom{1}}}}}\Biggr)\!
\Biggl(\frac{\epsilon^{l}_{\mu_{1}}({\bf k}_{1})}{\sqrt{2\omega^l_{{\bf k}_{1}}}}\Biggr)\!
\Biggl(\frac{\epsilon^{l}_{\mu_{2}}({\bf k}_{2})}{\sqrt{2\omega^l_{{\bf k}_{2}}}}\Biggr)\!
\,^{\ast}\Gamma^{\mu\mu_1\mu_2}(- k,- k_{1},- k_{2})\Bigr|_{\rm \,on-shell}.
\eqno{(6.9)}
\]
It should be noted that vertex functions (6.8) and (6.9) describe essentially different processes. By way of example, we consider the process described by the second diagram in Fig\,\ref{fig1} ($s$ channel). In fact, it includes two scattering subprocesses. In the approximation considered here, the first subprocess can be described as follows (Fig.\,\ref{fig2}):
\begin{figure}[hbtp]
\centering
\includegraphics[height=4.7cm]{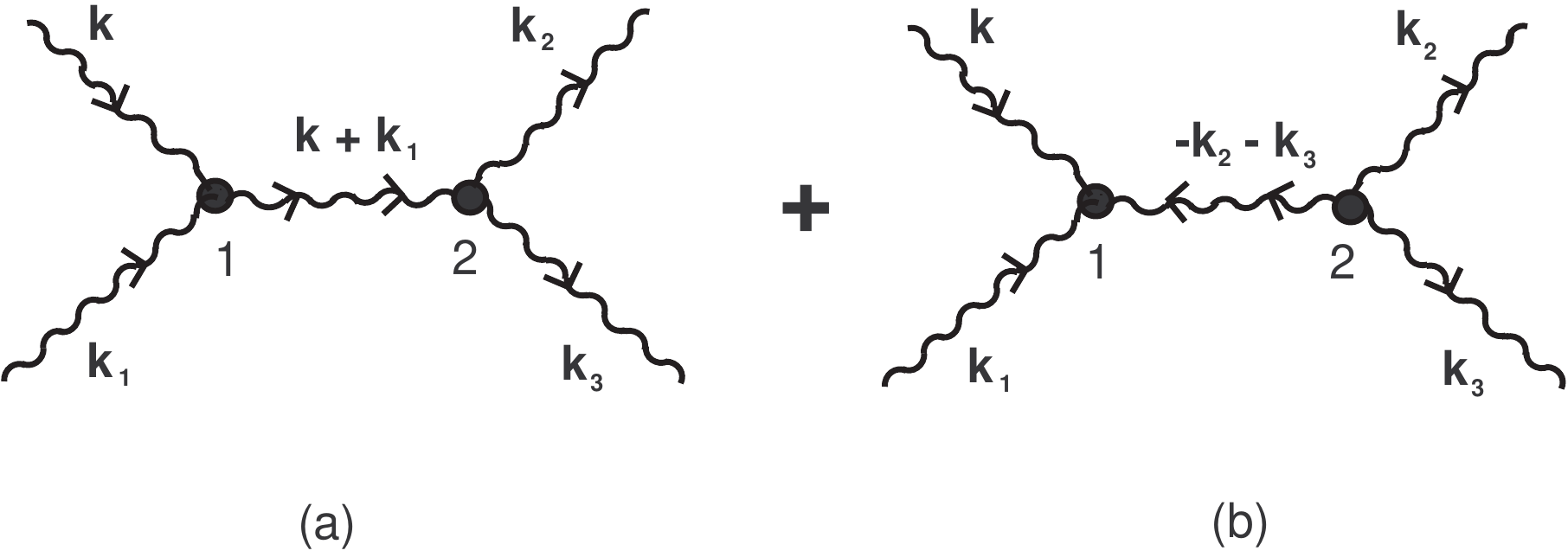}
\caption{Subprocesses of four-plasmon elastic scattering,
which are determined by processes of three-plasmon
decays and merging in the $s$ channel.}
\label{fig2}
\end{figure}
two plasmons with frequencies $\omega^{l}_{\bf k}$ and $\omega^{l}_{{\bf k}_{1}}$ and wavevectors ${\bf k}$ and ${\bf k}_{1}$ merge at vertex 1 into a single plasmon with frequency $\omega^{l}_{{\bf k} + {\bf k}_{1}}$ and wavevector ${\bf k} + {\bf k}_{1}$, which subsequently splits at vertex 2 into two plasmons with frequencies $\omega^{l}_{{\bf k}_{2}}$ and $\omega^{l}_{{\bf k}_{3}}$, and wavevectors ${\bf k}_{2}$ and ${\bf k}_{3}$ (Fig.\,\ref{fig2}(a)). In the classical Hamilton description, function  $1/(\omega^{l}_{{\bf k} + {\bf k}_{1}}\! - \omega^{l}_{{\bf k}} - \omega^{l}_{{\bf k}_{1}})$ plays the role of a propagator of an intermediate ``virtual'' state of collective longitudinal excitations, and the interaction at vertices 1 and 2 is defined in this case by function
$V_{{\bf k},\ {\bf k}_{1},\ {\bf k}_{2}}$ (6.8).\\
\indent The second subprocess is determined as follows (Fig.\,\ref{fig2}(b)): at vertex 2, a three-plasmon decay occurs -- two plasmons with frequencies $\omega^{l}_{{\bf k}_{2}}$ and $\omega^{l}_{{\bf k}_{3}}$ and wavevectors ${\bf k}_{2}$ and ${\bf k}_{3}$ pass to the system, while the third plasmon with frequency  $\omega^{l}_{{\bf k}_{2} + {\bf k}_{3}}$ and wavevector ${\bf k}_{2} + {\bf k}_{3}$ at vertex 1 merges with two plasmons with frequencies $\omega^{l}_{\bf k}$ and $\omega^{l}_{{\bf k}_{1}}$ and wavevectors ${\bf k}$ and ${\bf k}_{1}$, which arrive from the system. In this case, function $1/(\omega^{l}_{{\bf k} + {\bf k}_{1}}\! + \omega^{l}_{{\bf k}}+\omega^{l}_{{\bf k}_{1}})$ plays the role of a propagator. The interaction at vertices 1 and 2 is determined by function $U_{{\bf k},\ {\bf k}_{1},\ {\bf k}_{2}}$ (6.9).


\section{Conclusion}
\setcounter{equation}{0}
\label{section_10}

In this study, we have taken the first step in constructing the classical Hamiltonian formalism for describing processes of nonlinear interaction of soft gluon excitations in the Yang–Mills high-temperature field theory. We have constructed canonical transformation (2.10) in explicit form, which makes it possible to exclude third-order interaction Hamiltonian $\widehat{H}_{3}$ (2.6) and to define in this way new effective interaction Hamiltonian $\widetilde{\widehat{H}}_{4}$ (2.12) with gauge-invariant scattering amplitude $\widetilde{T}^{\ a\ a_{1}\ a_{2}\ a_{3}}_{{\bf k},\ {\bf k}_{1},\ {\bf k}_{2},\ {\bf k}_{3}}$. This interaction Hamiltonian determines a specific physical process, viz., elastic scattering of two colorless plasmons off each other. This scattering process dominates when the gauge field amplitude has order \cite{markov_2002}
\[
\vert A_{\mu}(x)\vert\sim\sqrt{g}T\,\;\mbox{and, accordingly}
\;N_{\bf k}^l\,\sim\,\displaystyle\frac{1}{g}\ \!,
\]
which in fact corresponds to the level of thermal fluctuations in a hot gluon plasma. For this value of the gauge field amplitude at $g\ll 1$, plasmon number density $N_{\bf k}^l$ is high, and the application of the purely classical description is justified. Moreover, the use of the linearized Boltzmann equation instead of the exact equation (4.4) is justified for colorless plasmons because the Planck distribution, relative to which the deviation $\delta N_{\bf k}^{l}$ of the plasmon number density is measured, is of order
\[
N_{\rm eq}^l({\bf k})\sim\displaystyle\frac{T}{\omega_{\bf k}^l}
\sim\displaystyle\frac{1}{g}\ \!.
\]
In this case, we can state that the theory of plasmon–plasmon interaction for small amplitudes of soft excitations is linear, and the nonlinear effects associated with nonequilibrium fluctuations $\delta N_{\bf k}^l$ of the plasmon number density can be treated as a perturbations.\\
\indent The situation changes qualitatively when the system is strongly excited, which can occur in collisions of ultrarelativistic heavy ions in experiments with the Large Hadron Collider. For high intensities of excitations in a gluon plasma, it is necessary to consider next terms in the expansion of $\widehat{H}_{int}$. Since nonlinear  excitation processes involving an odd number of plasmons are forbidden, we can in principle to get rid of all ``odd'' interaction Hamiltonians $\widehat{H}_{2n+1},\ n=1,2,\ldots\,$ by defining appropriately canonical transformations. In the limiting case of strong excitations, when 
\[
\vert A_{\mu}(x)\vert\sim T\,\;\mbox{and, accordingly,}
\;N_{\bf k}^l\,\sim\,\displaystyle\frac{1}{g^2\,},
\]
these canonical transformations contain an infinite number of terms of any order in creation operators $\hat{c}^{\dag\ \!\!a}_{{\bf k}}$ and annihilation ones $\hat{c}^{\ \!a}_{{\bf k}}$. In turn, this necessitates the inclusion of all higher-order plasmon elastic scattering processes ($3\rightarrow 3,\, 4\rightarrow 4,\,\ldots\,$) on the right-hand side of kinetic equation (4.4) since all these processes are of the same order in interaction constant $g$. Clearly, the procedure of linearization of the kinetic equation for plasmon number density $N_{\bf k}^l$ becomes inapplicable in the given case, and we arrive here at the truly nonlinear theory of interaction of soft gluon excitations in a plasma with a non-Abelian type of interaction.\\
\indent Thus, a nontrivial problem of constructing the explicit form of canonical transformations arises. These nonlinear canonical transformations must convert the original interaction Hamiltonian to a new effective form:
\[
\widehat{H}_{int}\longrightarrow\widetilde{\widehat{H}}_{int}=\widetilde{\widehat{H}}_{4}
+\widetilde{\widehat{H}}_{6}+\,\ldots\,+\widetilde{\widehat{H}}_{2n+2} + \,\ldots\,.
\]
However, the direct approach to determining the explicit form of required canonical transformations, which was used in this study, becomes ineffective in the attempt at the exclusion of even the next odd Hamiltonian $\widehat{H}_{5}$ because of extremely cumbersome calculations. For strongly excited states, when we are dealing with an infinite number of terms, a more adequate qualitatively new apparatus is required in the given situation (e.g., the introduction of a set of nonlocal canonical variables, which depend on an additional 3\hspace{0.02cm}D unit vector as proposed in \cite{metaxas_2001}. Another approach involves the use of relation
\[
A^{a}_{\mu}(k) = A^{(0)a}_{\mu}(k)
+\,^{\ast}\tilde{\cal D}_{\mu \nu}(k)\bigl\{\tilde{J}^{(2)a \nu}(A^{(0)},A^{(0)}) +
\tilde{J}^{(3)a\nu}(A^{(0)},A^{(0)},A^{(0)}) \,+\, \ldots\,\bigr\},
\eqno{(7.1)}
\]
where
$A^{\phantom{(0)}\!\!\!\!\!\!\!a}_{\mu}(k)$ and $A^{(0)a}_{\mu}(k)$ are the interacting and free gauge fields of the system, and the functions $\tilde{J}^{(n)a}_{\mu}(A^{(0)},A^{(0)},\,\ldots) $ are certain effective currents that are nonlinear functionals of the free field and are defined recurrently in the hard thermal loop approximation \cite{markov_2002}. The coefficient functions in $\tilde{J}^{(n)a}_{\mu}$ are effective amplitudes of type (5.3). As the interacting field, we must take expression (2.1), while as the free field, we must take the expression of form
$$
\hat{A}^{(0)a}_{\mu}(x) =\! \int\!\frac{d{\bf k}}{(2\pi)^{3}}\!\left(\frac{Z^{l}({\bf k})}
{2\omega^{l}_{{\bf k}}}\right)^{\!1/2}\!\!
\left\{\epsilon^{\ \! l}_{\mu}\ \!\hat{c}^{\ \!a}_{{\bf k}}\ \!e^{-i\hspace{0.02cm}{\bf k}\cdot{\bf x}}
\,+\,
\epsilon^{*\ \!}_{\mu}\ \!\hat{c}^{\dag\ \!\!a}_{{\bf k}}\ \!e^{i\hspace{0.02cm}{\bf k}\cdot{\bf x}}\right\}
$$
with operators $ \hat{c}^{\ \!a}_{{\bf k}}$ and $\hat{c}^{\dag\ \!a}_{{\bf k}}$ that appear on the right hand side of canonical transformations (2.10). Relation (7.1) in fact contains the required canonical transformation to any desired degree of accuracy if we use the appropriate approximations for propagators of type (6.6), (6.7) and vertex functions (6.8), (6.9), (5.8), etc. Relation (7.1) allows us to give a completely new interpretation of canonical transformations: transformations (2.10) determine a transition from noninteracting field $A^{(0)a}_{\mu}(k)$ to interacting field $A^{\phantom{(0)}\!\!\!\!\!\!\!a}_{\mu}(k)$, which takes into account all interaction effects in the medium. Analysis of this relationship requires separate consideration.


\section*{\bf Acknowledgments}

The research of D.M.G. and Yu.A.M. was supported by the Program for Improving Competitiveness of National Research Tomsk State University among the Leading World Scientific and Educational Centers. The work of D.M.G.
was also supported in part by the Russian Foundation for Basic Research (project no. 18-02-00149), San Paolo Research Foundation (FAPESP, project no. 2016/03319-6), and the National Science Council (CNPq).


\begin{appendices}
\numberwithin{equation}{section}

\section{Effective vertices and gluon propagator}

In this Appendix, we consider the explicit form of vertex functions and gluon propagator in the high-temperature hard thermal loop (HTL) approximation \cite{blaizot_2002, braaten_1990}.\\
\indent Effective three-gluon vertex
\[
\,^{\ast} \Gamma^{\mu \nu \rho}(k,k_1,k_2) \equiv
\Gamma^{\mu \nu \rho}(k,k_1,k_2) +
\delta \Gamma^{\mu \nu \rho}(k,k_1,k_2)
\eqno{({\rm A}.1)}
\]
is the sum of bare three-gluon vertex
\[
\Gamma^{\mu \nu \rho}(k,k_1,k_2) =
g^{\mu \nu} (k - k_1)^{\rho} + g^{\nu \rho} (k_1 - k_2)^{\mu} +
g^{\mu \rho} (k_2 - k)^{\nu}
\eqno{({\rm A}.2)}
\]
and the corresponding HTL correction
\[
\delta \Gamma^{\mu \nu \rho}(k,k_1,k_2) =
3\hspace{0.035cm}\omega^2_{\rm pl}\!\int\!\frac{d\hspace{0.035cm}\Omega}{4 \pi} \,
\frac{v^{\mu}v^{\nu}v^{\rho}}{v\cdot k + i\hspace{0.025cm}\epsilon} \,
\Biggl(\frac{\omega_2}{v\cdot k_2 - i\epsilon} -
\frac{\omega_1}{v\cdot k_1 - i\epsilon}\Biggr),
\eqno{({\rm A}.3)}
\]
where $v^{\mu} = (1,{\bf {\bf v}})$, $k  + k_1 + k_2 = 0$ and $d\hspace{0.035cm}\Omega$ is a differential solid angle. We consider below useful properties of the three-gluon HTL resummed vertex function for complex conjugation and permutation of momenta:
$$
\left(\!\,^{\ast}\Gamma_{\mu\mu_1\mu_2}(-k_1-k_2,k_1,k_2)\right)^{\ast} =
-\,^{\ast}\Gamma_{\mu\mu_1\mu_2}(k_1+k_2,-k_1,-k_2) 
\eqno{({\rm A}.4)}
$$
$$
= \!\,^{\ast}\Gamma_{\mu\mu_1\mu_2}(k_1+k_2,-k_2,-k_1).
$$
Further, the effective four-gluon vertex
\[
^{\ast} \Gamma^{\mu \nu \lambda \sigma}(k,k_1,k_2,k_3) \equiv
\Gamma^{\mu \nu \lambda \sigma}(k,k_1,k_2,k_3) +
\delta \Gamma^{\mu \nu \lambda \sigma}(k,k_1,k_2,k_3)
\eqno{({\rm A}.5)}
\]
is the sum of bare four-gluon vertex
\[
\Gamma^{\mu \nu \lambda \sigma} =
2\hspace{0.03cm}g^{\mu \nu}g^{\lambda \sigma} - g^{\mu \sigma}g^{\nu \lambda} -
g^{\mu \lambda}g^{\sigma \nu}
\eqno{({\rm A}.6)}
\]
and the corresponding HTL correction
\[
\delta \Gamma^{\mu \nu \lambda \sigma}(k, k_1, k_2,k_3) 
= 
3\hspace{0.035cm}\omega^2_{\rm pl}\!\int\!\frac{d\hspace{0.035cm}\Omega}{4 \pi} \, \frac{v^{\mu}v^{\nu}v^{\lambda}v^{\sigma}}{v\cdot k + i\hspace{0.025cm}\epsilon}
\eqno{({\rm A}.7)}
\]
\[
\times\Biggl[
\,\frac{1}{v\cdot (k + k_1) + i\epsilon}\, 
\Biggl(\frac{\omega_{2}}{v\cdot k_2 - i \epsilon} - \frac{\omega_3}{v\cdot k_3 - i \epsilon} \Biggr)
- \frac{1}{v\cdot (k + k_3) + i \epsilon}\,
\Biggl(\frac{\omega_{1}}{v\cdot k_1 - i \epsilon} - \frac{\omega_2}{v\cdot k_2 - i \epsilon} \Biggr)\Biggr].
\]
Finally, the expression
\[
^{\ast}\widetilde{\cal D}_{\mu \nu}(k) = 
- P_{\mu \nu}(k) \,^{\ast}\!\Delta^t(k) - \widetilde{Q}_{\mu \nu}(k) \,^{\ast}\!\Delta^l(k)
- \xi_{0}\ \!\frac{k^{2}}{(k\cdot u)^{2}}\ \!D_{\mu \nu}(k)
\eqno{({\rm A}.8)}
\]
is a gluon (retarded) propagator in the $A_0$\hspace{0.02cm}-\hspace{0.02cm}gauge, which is modified by effects of the medium. Here, ``scalar'' transverse and longitudinal propagators have form
\[
\hspace{-1cm}\,^{\ast}\!\Delta^{t}(k) = \frac{1}{k^2 - \Pi^{t}(k)},
\qquad\quad\;
\,^{\ast}\!\Delta^{l}(k) = \frac{1}{k^2 - \Pi^{l}(k)},
\eqno{({\rm A}.9)}
\]
where
\[
\Pi^{\hspace{0.025cm} t}(k) = \frac{1}{2}\Pi^{\mu\nu}(k) P_{\mu\nu}(k),
\qquad
\Pi^{\hspace{0.025cm} l}(k) = \Pi^{\mu\nu}(k) \widetilde{Q}_{\mu\nu}(k).
\hspace{0.2cm}
\]
The polarization tensor $\Pi_{\mu \nu}(k)$ in the HTL approximation has form
\[
\Pi^{\mu \nu}(k) = 3\hspace{0.035cm}\omega_{\rm pl}^{2}
\left( u^{\mu}u^{\nu} - \omega\!\int\!\frac{d\hspace{0.035cm}\Omega}{4 \pi}
\,\frac{v^{\mu}v^{\nu}}{v\cdot k + i \epsilon} \right)
\]
and the longitudinal and transverse projectors are defined by the expressions
\[
\widetilde{Q}_{\mu \nu}(k) =
\frac{\tilde{u}_{\mu}(k) \tilde{u}_{\nu}(k)}{\bar{u}^2(k)}\, ,
\eqno{({\rm A}.10)}
\]
\[
P_{\mu\nu}(k) = g_{\mu\nu} - u_{\mu}u_{\nu}
- \widetilde{Q}_{\mu \nu}(k)\,\frac{(k\cdot u)^{2}}{k^{2}}\, ,
\]
respectively, where Lorentz-covariant four-vector $\tilde{u}_{\mu}(k)$ is defined by the formula (5.5).

\end{appendices}

\end{document}